\documentclass[a4paper,12pt]{article}

\usepackage{amsmath}
\usepackage{amssymb}
\usepackage{slashed}
\usepackage{amsthm}
\usepackage{xcolor}

\numberwithin{equation}{section}
\def\appendix#1{\addtocounter{section}{1}\setcounter{equation}{0}
\renewcommand{\thesection}{\Alph{section}}
\section*{Appendix \thesection\protect\indent \parbox[t]{11.15cm}{#1}}
\addcontentsline{toc}{section}{Appendix \thesection\ \ \ #1}}

\newcommand{\bea}{\begin{eqnarray}}
\newcommand{\eea}{\end{eqnarray}}

\begin{document}

\begin{titlepage}
\begin{center}

\vspace*{-1.0cm}

\hfill  DMUS-MP-21-07
\\
\vspace{2.0cm}

\renewcommand{\thefootnote}{\fnsymbol{footnote}}
{\Large{\bf $N=4$ Near-Horizon Geometries in $D=11$ Supergravity }}
\vskip1cm
\vskip 1.3cm
D. Farotti and J. Gutowski
\vskip 1cm
{\small{\it
Department of Mathematics,
University of Surrey \\
Guildford, GU2 7XH, UK.}\\
\texttt{d.farotti@surrey.ac.uk, j.gutowski@surrey.ac.uk}}

\end{center}
\bigskip
\begin{center}
{\bf Abstract}
\end{center}

Extreme near-horizon geometries in $D=11$ supergravity preserving four supersymmetries are classified. It is shown that the Killing spinors fall 
into three possible orbits, corresponding to pairs of spinors defined on the spatial cross-sections of the horizon which have isotropy groups $SU(3)$, $G_2$, or $SU(4)$. In each case, the conditions on the geometry and the 4-form flux are determined. The integrability conditions obtained from the Killing spinor equations are also investigated.

\end{titlepage}

\section{Introduction}

Black hole uniqueness is of particular interest in the context of string theory.
The strongest black hole uniqueness theorems are formulated
for asymptotically flat  solutions in four dimensions \cite{israel, carter, hawking, robinson1, israel2, mazur} . However, it is notable that the inclusion of a (negative) cosmological constant weakens these types of uniqueness theorems, even in four dimensions
\cite{Gnecchi:2013mja, Vanzo:1997gw}. The situation is furthermore complicated in relation to black holes in string theory, which frequently arise 
in higher dimensional supergravities. In this case the notion of black hole uniqueness is also weaker. The first signal of this was the discovery of asymptotically flat black rings (both with, and without, supersymmetry) \cite{Emparan:2001wn, Elvang:2004rt, Elvang:2004ds}, for which the 
event horizon spatial topology is $S^2 \times S^1$. Examples of black rings can be constructed in such a way as to have the same asymptotic charges as black holes with $S^3$ event horizon topology.
Additional evidence of non-uniqueness has also been more recently discovered in \cite{Breunholder:2018roc, Breunholder:2017ubu}, where a very large class of extreme, supersymmetric, black hole solutions has been found. These 
are both asymptotically flat, and have near-horizon geometries which include the geometry of the near-horizon BMPV solution \cite{Breckenridge:1996is} as a possibility, but which in addition have non-trivial topology located outside of the horizon.
    The charges associated with such solutions include the mass, angular momenta and electric charge which are evaluated at infinity; but in addition to these, there are also magnetic charges supported on certain, but arbitrarily many, 2-cycles.
The thermodynamic properties of such solutions have also been considered in \cite{Kunduri:2018qqt}.
 So, there is a large family
of black hole geometries in five dimensions, which includes the BMPV black hole as a special case,
but for which there is, a priori, no upper bound on the number of independent parameters which can
be used to construct the geometries in this family. This is in marked contrast to the classical uniqueness theorems in four dimensions.

Nevertheless, in spite of this increased complexity,
there are uniqueness results established when there are sufficiently many 
commuting rotational isometries \cite{Harmark:2004rm, Hollands:2008fm, Alaee:2019qhj, Hollands:2007aj}. Uniqueness theorems for static and asymptotically flat higher dimensional black hole solutions have also been established \cite{Gibbons:2002bh}.
        It is however notable that, even given such a rich structure of higher dimensional black
hole solutions, there are currently no known analytic examples of
        asymptotically de-Sitter, or anti-de-Sitter, black rings in higher dimensions
           with regular horizons, and there are also non-existence theorems for such objects in certain theories \cite{Grover:2013hja, Kunduri:2006uh, Lucietti:2018osk, Khuri:2017zqg}. It would appear that additional matter fields are required to be included for such solutions to exist.

Motivated by this, it is natural to investigate the issue of black hole uniqueness for extreme and supersymmetric black holes in eleven-dimensional supergravity. We remark that, in contrast to a large number of four and 
five-dimensional supergravity theories, black hole solutions in D=11 supergravity need not be extreme even if supersymmetry is imposed. This is because the algebraic identities satisfied by the gauge-invariant spinor bilinears differ between D=11 supergravity and four and five-dimensional theories. Hence, in this work, extremality will be imposed as an extra assumption in addition to supersymmetry, which enables the near-horizon limit of the black hole to be considered. 

 Although higher dimensional black holes are not determined uniquely by their near-horizon geometry, 
investigating the geometric structures associated with such near-horizon solutions is a necessary first step in establishing a systematic classification of these types of black holes. As we have mentioned above, even in 
the presence of supersymmetry, black hole uniqueness breaks down
in higher dimensions. However, the conditions on the geometry following from supersymmetry can be used to put constraints on the possible solutions, and in some cases to eliminate certain geometries entirely \cite{Grover:2013hja}. More generally near-horizon geometries typically admit additional symmetries when compared to the bulk black hole geometry, and this makes the analysis  more tractable.

In terms of D=11 supergravity, there has been some significant progress 
in understanding the properties of supersymmetric near-horizon geometries. In particular, in \cite{mhor}, it was proven that all supersymmetric near-horizon geometries in D=11 supergravity must admit an even number of supersymmetries, and furthermore all such solutions admit a $SL(2, \mathbb{R})$ symmetry algebra. The geometric structure of near-horizon solutions admitting $N=2$ supersymmetry was classified in \cite{mhor, Gutowski:2012eq}. Following on from this, analysis of the superalgebras associated with supersymmetric near-horizon geometries \cite{super1, super2} was used to further elucidate the nature of the symmetry algebras of solutions with $N>2$ 
supersymmetry. In conjunction with the homogeneity theorem in \cite{Figueroa-OFarrill:2012kws}, 
together with a classification of all compact 9-dimensional Riemannian manifolds in 
\cite{klaus}, it was shown that there are no supersymmetric near horizon geometries with
$16<N<32$ supersymmetries. Any maximally supersymmetric $N=32$ near-horizon geometry must furthermore be one of the maximally supersymmetric geometries determined in \cite{Figueroa-OFarrill:2002ecq}. 

It therefore remains to consider supersymmetric near horizon geometries with an even number $4 \leq N \leq 16$  of supersymmetries, and in this work we shall consider the case of $N=4$. We remark that the cases of $6 \leq N \leq 16$ supersymmetries are not readily amenable to analysis via the techniques used in this paper. This is because, in order to determine 
the conditions obtained from the Killing spinor equations, we first explicitly integrate these equations along two of the light-cone directions which are generic to all near-horizon solutions. The remaining content of the Killing spinor equations reduces to finding spinors whose components depend only on the 9 remaining co-ordinates associated with the 9-dimensional spatial cross-section of the event horizon, which are parallel with respect to certain 9-dimensional supercovariant derivatives including contributions from the 4-form flux.  To extract conditions on the geometry and fluxes from requiring the existence of such parallel spinors, $Spin(9)$ 
gauge transformations are used to simplify the form of these spinors.
    Such simplifications can be implemented in the cases of $N=2$ and $N=4$ supersymmetry. However, the majority of this gauge freedom is used 
up in simplifying the spinors for the $N=4$ case, and little further useful simplification can be made to the extra spinors for $N=6$ solutions. Nevertheless, it is clear that for $6 \leq N \leq 16$, the additional supersymmetry will impose further conditions on the geometry, and it would be interesting to consider these cases in more detail in future work.


In this paper we determine the conditions on the geometry of the 9-dimensional spatial horizon section, and also the 4-form flux, of extreme supersymmetric black hole near-horizon geometries 
preserving $N=4$ supersymmetry in $D=11$ supergravity. To this end, we shall solve the horizon Killing spinor equations (KSE) using spinorial geometry techniques \cite{Gillard:2004xq, lawson}. This method has been used extensively to construct classifications of supergravity solutions (e.g. \cite{Gran:2005wn}), as well as to investigate properties of highly 
supersymmetric solutions \cite{Gran:2006cn} and black hole solutions \cite{Gutowski:2012eq}; see also \cite{revhor} and references therein for a comprehensive list of applications.
The key features of the spinorial geometry approach are firstly that appropriate gauge transformations are utilized to write explicitly the spinors in the simplest possible canonical form. Using this, the conditions obtained from the KSE can then be evaluated directly. This approach to analysing the KSE is particularly useful when investigating the 
properties of solutions with more than the minimal possible amount of supersymmetry, as it is not necessary to consider the rather detailed analysis of Fierz identities which would otherwise be required.

 We apply these techniques to the $N=4$ near-horizon geometries by determining explicitly the simple canonical forms for the spinors. In particular, the conditions required for $N=4$ supersymmetry are equivalent to requiring the existence of two Majorana spinors whose components depend only on the horizon section co-ordinates. These two spinors must be positive chirality, with respect to a certain light-cone chirality projection, and must also be 
orthonormal as a consequence of the global analysis of \cite{mhor}. Using 
appropriate gauge transformations, we find that there are three different 
stabilizers, namely $SU(3)$, $G_2$ or $SU(4)$, for the two positive chirality spinors. The three cases are treated separately. In each case, using 
the spinorial geometry techniques, we find the constraints on the geometry and we express the fluxes in terms of the geometry in a covariant fashion using the gauge-invariant bilinears which are constructed from the spinors.

The paper is organized as follows. In section 2 we present the near-horizon fields of $D=11$ supergravity and write the field equations, the Bianchi identities and the Killing spinor equations in terms of these near-horizon fields. In particular, the conditions required for supersymmetry reduce to finding spinors which depend only on the co-ordinates of the spatial horizon section, and are parallel with respect to certain supercovariant derivatives defined on the horizon sections. We also review some global and Killing superalgebra analysis. In section 3 we present the solution of the Killing spinor equations of near-horizon geometries preserving $N=2$ supersymmetries, fully expressing it in terms of the $Spin(7)$ gauge-invariant bilinears constructed out of the spinors. In section 4 we solve the Killing spinor equations of near-horizon geometries preserving $N=4$ supersymmetries, splitting the analysis in three distinct cases, $SU(3)$, $G_2$ or $SU(4)$, corresponding to the isotropy groups of the spinors associated with the $N=4$ solutions. In section 5 we analyse the integrability conditions of near-horizon geometries in $D=11$ supergravity. Using a purely local argument, we shall show that all the components of the Einstein equations are implied by the 11-dimensional KSE, the 4-form field equations and the Bianchi identites. The paper concludes with four appendices, where we have gathered some of the lengthier formulae, namely the $N=4$ linear systems associated to the KSE, the expression of the 4-form in terms of the $SU(3)$ gauge-invariant bilinears, and various $G_2$ and $Spin(7)$ identities.

\section{Supersymmetric Near-Horizon Geometries}

In this section, we collate and summarize the key results obtained from
{\cite{smhor}}, {\cite{mhor}}, {\cite{Gutowski:2012eq}}, {\cite{super1}}, 
{\cite{super2}}, which we shall utilize 
in considering the analysis of near-horizon geometries.

Using Gaussian null coordinates \cite{isen, gnull},  the near horizon metric and the 4-form flux \footnote{Let $\omega$ be a k-form, then
$d_h\omega:= d\omega-h\wedge \omega$.} of $D=11$ supergravity can be written \cite{mhor, Gutowski:2012eq, smhor} in the following form
\begin{eqnarray}
ds^2 &=& 2 {\bf{e}}^+ {\bf{e}}^- + \delta_{ij} {\bf{e}}^i {\bf{e}}^j=2 du (dr + r h - {1 \over 2} r^2 \Delta du)+ ds^2({\cal S})~,
\cr
F &=& {\bf{e}}^+ \wedge {\bf{e}}^- \wedge Y
+ r {\bf{e}}^+ \wedge d_h Y + X~,
\label{mhm}
\end{eqnarray}
where the vielbein $\{{\bf{e}}^{+}, {\bf{e}}^{-}, {\bf{e}}^{i}\}$, with $i=1, 2, 3, 4, 6, 7, 8, 9, \sharp$ is given by
\begin{eqnarray}
{\bf{e}}^+ = du~,~~~{\bf{e}}^- = dr + r h - {1 \over 2} r^2 \Delta du~,~~~
{\bf{e}}^i = e^i{}_J dy^J~,~~~g_{IJ}=\delta_{ij} e^i{}_I e^j{}_J~,
\label{nhbasis}
\end{eqnarray}
and
\begin{eqnarray}
ds^2({\cal S})=\delta_{ij} {\bf{e}}^i{\bf{e}}^j~,
\end{eqnarray}
is the metric of the horizon spatial cross-section ${\cal S}$ given by $r=u=0$. In these expressions,  $h=h_i {\bf{e}}^i$, $\Delta$, ${\bf{e}}^i$, $Y$ and 
$X$ depend only on  the co-ordinates $y$
of ${\cal S}$. Furthermore, we assume that  ${\cal S}$ is a compact 9-dimensional manifold without boundary, and that $\Delta$,  $h$, $Y$ and $X$ are globally defined and smooth 0-, 1-, 2- and 4-forms on ${\cal S}$, respectively.

\subsection{Bianchi identities and field equations}

The Bianchi identities and field equations of $D=11$ supergravity 
\cite{julia} can be decomposed along the lightcone directions and those of the horizon section ${\cal S}$. For the Bianchi identities $dF=0$, such a decomposition yields
\begin{eqnarray}
\label{clos}
dX=0 \ .
\end{eqnarray}
The field equation of the 4-form flux $F$ reads
\begin{eqnarray}
d \star_{11}  F -{1 \over 2} F \wedge F=0~,
\label{Maxwell}
\end{eqnarray}
where $\star_{11}$ is the Hodge star on 11-dimensional spacetime. Equation \eqref{Maxwell}  can be written as
\begin{eqnarray}
\label{geq1}
-\star_9 d_hY  - h \wedge \star_9 X + d \star_9 X = Y \wedge X~,
\end{eqnarray}
and
\begin{eqnarray}
\label{geq2}
-d \star_{9} Y = {1 \over 2} X \wedge X~,
\end{eqnarray}
where $\star_{9}$ denotes the Hodge dual on ${\cal S}$. The volume form  is chosen as $\epsilon_{11}= {\bf{e}}^+ \wedge {\bf{e}}^- \wedge \epsilon_{{\cal{S}}}$, where  $\epsilon_{{\cal{S}}}$
is the volume form of ${\cal S}$.

The Einstein equation is
\begin{eqnarray}
R_{MN} = {1 \over 12} F_{M L_1 L_2 L_3} F_N{}^{L_1 L_2 L_3}
-{1 \over 144} g_{MN} F_{L_1 L_2 L_3 L_4}F^{L_1 L_2 L_3 L_4} \ ,
\label{Einstein}
\end{eqnarray}
for which the independent components are the $ij$ components
\begin{eqnarray}
\label{ein1}
{\tilde {R}}_{ij} + {\tilde{\nabla}}_{(i} h_{j)} -{1 \over 2} h_i h_j &=& -{1 \over 2} Y_{i \ell} Y_j{}^\ell
+{1 \over 12} X_{i \ell_1 \ell_2 \ell_3} X_j{}^{\ell_1 \ell_2 \ell_3}
\nonumber \\
&+& \delta_{ij} \bigg( {1 \over 12} Y_{\ell_1 \ell_2} Y^{\ell_1 \ell_2}
-{1 \over 144} X_{\ell_1 \ell_2 \ell_3 \ell_4} X^{\ell_1 \ell_2 \ell_3 \ell_4} \bigg)~,
\end{eqnarray}
 and the $+-$ component
\begin{eqnarray}
\label{einpm}
{\tilde{\nabla}}^i h_i = 2 \Delta + h^2 -{1 \over 3} Y_{\ell_1 \ell_2} Y^{\ell_1 \ell_2}
-{1 \over 72} X_{\ell_1 \ell_2 \ell_3 \ell_4} X^{\ell_1 \ell_2 \ell_3 \ell_4}~.
\end{eqnarray}
Here ${\tilde{\nabla}}$ is the Levi-Civita connection of the metric $ds^2({\cal S})$ and ${\tilde{R}}_{ij}$ is the Ricci tensor of ${\cal{S}}$. The $++$ and $+i$ components of the Einstein equation hold as a consequence of
({\ref{clos}}), the 3-form field equations ({\ref{geq1}}) and ({\ref{geq2}}) and
the components of the Einstein equation in ({\ref{ein1}}) and  ({\ref{einpm}}).
The $--$ and the $-i$ components of the Einstein equations are also satisfied
automatically for all near-horizon solutions ({\ref{mhm}}).
Thus, the conditions on $ds^2({\cal S})$,  $\Delta$, $h$, $Y$ and $X$ simplify to
({\ref{clos}}), ({\ref{geq1}}), ({\ref{geq2}}), ({\ref{ein1}}) and ({\ref{einpm}}).

\subsection{Independent Killing Spinor Equations}

The Killing spinor equations (KSE) of 11-dimensional supergravity \cite{julia} are
\begin{eqnarray}
\nabla_M \epsilon
+\bigg(-{1 \over 288} \Gamma_M{}^{L_1 L_2 L_3 L_4} F_{L_1 L_2 L_3 L_4}
+{1 \over 36} F_{M L_1 L_2 L_3} \Gamma^{L_1 L_2 L_3} \bigg) \epsilon =0~,
\nonumber \\
\label{KSE11dimensional}
\end{eqnarray}
where $\nabla$ is the Levi-Civita connection. The KSE can be integrated along the two lightcone
directions, as all the bosonic fields are independent of the $u$ co-ordinate, and the 
dependence on the $r$ co-ordinate is known explicitly. The remaining content of
the KSE can then be further decomposed along the directions corresponding to the spatial
horizon section ${\cal{S}}$. To begin, we write the Killing spinor $\epsilon$ as \cite{Gutowski:2012eq}

\begin{eqnarray}
\epsilon = \epsilon_+ + \epsilon_-~,~~~\Gamma_\pm \epsilon_\pm =0~.
\label{dec}
\end{eqnarray}

Then after some computation, see \cite{mhor, Gutowski:2012eq, smhor}, we find that
\begin{eqnarray}
\label{ksp1}
\epsilon_+ = \eta_+, \qquad \epsilon_- = \eta_- + r \Gamma_-
\Theta_+\eta_+~,
\label{ksp3}
\end{eqnarray}
and
\begin{eqnarray}
\label{ksp2}
\eta_+ = \phi_+ + u \Gamma_+ \Theta_- \phi_- , \qquad \eta_- = \phi_-~,
\end{eqnarray}
where $\phi_\pm = \phi_\pm (y)$ do not depend on $r$ or $u$ and
\begin{eqnarray}
\Theta_\pm=\bigg({1 \over 4} h_i \Gamma^i +{1 \over 288} X_{\ell_1 \ell_2 \ell_3 \ell_4}
\Gamma^{\ell_1 \ell_2 \ell_3 \ell_4} \pm {1 \over 12} Y_{\ell_1 \ell_2} \Gamma^{\ell_1 \ell_2} \bigg)~.
\end{eqnarray}

Using the field equations and Bianchi identities, the
independent KSE are
\begin{eqnarray}
\label{ind}
\nabla_i^{(\pm)}\phi_ \pm \equiv {\tilde{\nabla}}_i \phi_\pm + \Psi^{(\pm)}_i \phi_\pm =0~,
\end{eqnarray}
where
\begin{eqnarray}
\Psi^{(\pm)}_i &=& \mp{1 \over 4} h_i -{1 \over 288} \Gamma_i{}^{\ell_1 
\ell_2 \ell_3 \ell_4}
X_{\ell_1 \ell_2 \ell_3 \ell_4} +{1 \over 36} X_{i \ell_1 \ell_2 \ell_3} \Gamma^{\ell_1 \ell_2 \ell_3}
\nonumber \\
&\pm&{1 \over 24} \Gamma_i{}^{\ell_1 \ell_2} Y_{\ell_1 \ell_2} \mp{1 \over 6} Y_{ij} \Gamma^j~.
\end{eqnarray}

\subsection{Global and Superalgebra Analysis of Supersymmetry}

For each of the ``horizon gravitino KSEs'' on ${\cal S}$ given in ({\ref{ind}}), one can associate a ``horizon Dirac equation''  as
\begin{eqnarray}
\label{dirac1}
{\cal{D}}^{(\pm)}\phi_\pm \equiv \Gamma^i \nabla_i^{(\pm)}\phi_ \pm = \Gamma^i {\tilde{\nabla}}_i \phi_\pm + \Psi^{(\pm)} \phi_\pm =0 \ ,
\end{eqnarray}
where
\begin{eqnarray}
\Psi^{(\pm)} = \Gamma^i \Psi^{(\pm)}_i = \mp{1 \over 4} h_\ell \Gamma^\ell +{1 \over 96} X_{\ell_1 \ell_2 \ell_3 \ell_4}
\Gamma^{\ell_1 \ell_2 \ell_3 \ell_4} \pm{1 \over 8} Y_{\ell_1 \ell_2} \Gamma^{\ell_1 \ell_2}~.
\end{eqnarray}
These Dirac equations,  in addition to the Levi-Civita connection, also depend on the fluxes of the supergravity theory restricted on the horizon section ${\cal S}$, and the properties of the horizon Dirac equations play an important role in counting the number of supersymmetries preserved by the solutions. In particular, generalized Lichnerowicz Theorems established in \cite{mhor} (see also
analogous results for type II supergravity theories in \cite{iibhor,iiahora,iiahorb, revhor}) imply that
\begin{eqnarray}
\nabla^{(\pm)}_i \phi_\pm=0\Longleftrightarrow {\cal D}^{(\pm)} \phi_\pm=0~,
\end{eqnarray}
and furthermore, via an index theory argument, we have
\begin{eqnarray}
{\rm dim} \, {\rm ker} {\cal{D}}^{(+)} = {\rm dim}\, {\rm ker} {\cal{D}}^{(-)}~.
\end{eqnarray}

The global analysis which is used to establish the generalized Lichnerowicz theorems, also implies, by a maximum principle argument, that $\parallel \phi_+ \parallel^2$ is constant, and henceforth we shall take all such positive chirality spinors to satisfy $\parallel \phi_+ \parallel^2=1$. 
The number of positive chirality spinors which are parallel with respect to 
$\nabla_i^{(+)}$ is equal to the number of negative chirality spinors which are parallel with
respect to $\nabla_i^{(-)}$, and for a supersymmetric near-horizon geometry preserving $2k$ supersymmetries, we take $\phi^{r}_{\pm}$ spinors for $r=1, \dots k$ with, without loss of
generality, $\langle \phi^{r}_+, \phi^{s}_+ \rangle = \delta^{rs}$, where $\langle, \rangle$ is a Spin(9)-invariant Dirac inner product.

There is furthermore an algebraic relation which can be used to construct 
$\phi^r_+$ spinors from $\phi^r_-$ spinors which are parallel with respect to $\nabla_i^{(\pm)}$ via
 \begin{eqnarray}
\label{extragen}
\phi^{r}_+ = \Gamma_+ \Theta_- \phi^{r}_-
\end{eqnarray}
and in particular, it has been shown that if $\phi^{r}_- \neq 0$ then $\phi^{r}_+$ defined as above is non-vanishing. A further algebraic relation, based on a superalgebra computation \cite{super1, super2}, has also been constructed which relates $\phi^r_-$ to $\phi^r_+$ via
\bea
2\phi_-^{t}-2 \parallel \phi_-^{t} \parallel^2 \Gamma_-\Theta_+\phi_+^{t}+\slashed{\check{V}}\Gamma_-\phi_+^{t}=0\,.
\label{implicit}
\eea
where ${\check{V}}$ is obtained from the horizon section isometries
\begin{eqnarray}
\label{extraiso}
V^{rs}_i\equiv  \langle \Gamma_+ \phi_-^{r} , \Gamma_i \phi_+^{s} \rangle
\end{eqnarray}
via
\begin{eqnarray}
\label{extraiso}
V^{rs} =  \delta^{rs} {\check{V}} + Z^{rs} ~,
\end{eqnarray}
where $Z^{rs}$ is antisymmetric in the indices $r,s$. Moreover, the negative 
chirality spinors $\phi_-^r$ satisfy
\begin{eqnarray}
\langle \phi_-^r, \phi_-^s \rangle = \delta^{rs} \parallel \phi_-^{r} \parallel^2 \ ,
\end{eqnarray}
and the positive chirality spinors $\phi_+^r$ satisfy
\begin{eqnarray}
4 \langle \Theta_+ \phi_+^r , \Theta_+ \phi_+^s \rangle = \delta^{rs} \Delta~,
\qquad \langle \phi_+^{(r}, \Gamma_i \Theta_+ \phi_+^{s)} \rangle =0 \ .
\end{eqnarray} 
The isometries $V_i^{rs}$ on ${\cal{S}}$ are obtained from $D=11$ spacetime
isometries as set out in \cite{super1, super2}, and both the $D=11$
metric and 4-form $F$ are invariant with respect to these isometries
\cite{gpk}. Consequently, the isometries $\check{V}$ and $Z^{rs}$ 
leave invariant all of the near-horizon data:
\begin{eqnarray}
\tilde\nabla_{(i} \check{V}_{j)}=0~,~~~\tilde {\cal L}_{\check{V}}h=0 
~,~~~\tilde {\cal L}_{\check{V}}\Delta=0~,~~~\tilde {\cal L}_{\check{V}} Y=0~,~~~\tilde {\cal L}_{\check{V}} X=0~,
\end{eqnarray}
and similarly
\begin{eqnarray}
\tilde\nabla_{(i} Z^{rs}_{j)}=0~,~~~\tilde {\cal L}_{Z^{rs}}h=0 ~,~~~\tilde {\cal L}_{Z^{rs}}\Delta=0~,~~~\tilde {\cal L}_{Z^{rs}} Y=0~,~~~\tilde {\cal L}_{Z^{rs}} X=0~.
\end{eqnarray}
Furthermore, $\check{V}$ and $Z^{rs}$  commute
\begin{eqnarray}
[\check{V}, Z^{rs}]=0 \ .
\end{eqnarray}

The condition ({\ref{implicit}}) does not however determine the $\phi^r_-$ spinors directly from the $\phi^r_+$ spinors because $\check{V}$ depends implicitly on $\phi^r_-$ via
({\ref{extraiso}}). The vector field $\check{V}$ may vanish, in such a case the
near-horizon geometry is static, and is a warped product $AdS_2 \times_w {\cal{S}}$
\cite{mhor}. Generically, a non-static $N=4$ solution has non-vanishing
commuting isometries $\check{V}$ and $Z=Z^{12}$, and therefore admits a 
$U(1) \times U(1)$ symmetry on the horizon spatial section ${\cal{S}}$. However, there are also special cases when this symmetry is reduced. Additional geometric conditions can be obtained from a detailed analysis of the Killing spinor equations, which we shall undertake in the following sections. We remark that it will be most straightforward to consider the conditions obtained by solving
\begin{eqnarray}
\nabla_i^{(+)}\phi^r_+ =0~,
\end{eqnarray}
where $r=1$ for $N=2$ solutions, and $r=1,2$ for $N=4$ solutions. 

This is because the constant norm condition  $\langle \phi^{r}_+, \phi^{s}_+ \rangle = \delta^{rs}$, combined with a judicious application of $Spin(9)$ gauge transformations, allows one to find very simple canonical forms for the 
$\phi_+^r$ spinors, and consequently the linear systems obtained from the 
KSE
are rendered more tractable. Having solved the KSE for the positive chirality
spinors, the global analysis guarantees that there must exist negative 
chirality spinors $\phi_-^r$ which are parallel with respect to $\nabla^{(-)}$.

\section{$N=2$ Solutions}
In this section we revisit near-horizon geometries preserving exactly $N=2$ supersymmetries. This has already been discussed in \cite{Gutowski:2012eq} working in a particular gauge for the $\phi_-^1$ spinor. However, our analysis here does not involve that gauge fixing procedure, as we wish to retain sufficient gauge freedom to simplify the additional positive chirality spinor which we will consider in the later analysis of $N=4$ solutions. It is convenient to solve the KSE in the positive chirality sector
\bea
\nabla_{i}^{(+)}\phi^1_{+}=0\,,
\label{KSEN=2}
\eea
since, as we have mentioned in the previous section, $\phi_+^1$ has constant norm, which enables
us to write this spinor in a particularly simple canonical form. In particular, using the $Spin(9)$ gauge invariance of the KSEs (\ref{KSEN=2}), 
namely
\bea
\phi^1_{+}\to e^{f^{ij}\Gamma_{ij}}\phi^1_{+}\,,
\label{spin9}
\eea
where $\Gamma_{ij}$ are the generators of $Spin(9)$, the spinor $\phi_{+}^1$ can be written in the following particularly simple canonical form
\bea
\phi_{+}^{\,1}=\frac{1}{\sqrt{2}}(1+e_{1234})~,
\label{phi1}
\eea
which is stabilized by a $Spin(7)$ isotropy subgroup of $Spin(9)$ \cite{smhor}.

It will be particularly useful to demonstrate explicitly how this can be achieved, as it
will illustrate how a similar analysis can be done to determine the common isotropy
group of two spinors $\{ \phi_+^1, \phi_+^2 \}$ in the next section, which is the primary
focus of this work. So, in order to show (\ref{phi1}), let us start by writing down the general expression for a positive (lightcone) chirality Majorana spinor \cite{system11}
\bea
\phi_+^1=w1+\bar{w}e_{1234}+\psi^{\alpha}e_{\alpha}-\frac{1}{3!}(\star\bar{\psi})^{\alpha_1 \alpha_2 \alpha_3}e_{\alpha_1 \alpha_2 \alpha_3}+\frac{1}{2}(B^{\alpha_1 \alpha_2}-(\star\bar{B})^{\alpha_2 \alpha_2})e_{\alpha_1 \alpha_2}\,,
\label{Majorana1}
\eea
where in (\ref{Majorana1}) $\alpha, \beta =1,2,3,4$, $\star$ is the Hodge dual on $\mathbb{R}^4$ and $w$, $\psi$ and $B$ are functions of the horizon coordinates $y$. Defining $A^{\alpha \beta}:=B^{\alpha \beta}-(\star\bar{B})^{\alpha \beta}$, we can rewrite (\ref{Majorana1}) as follows
\bea
\phi^1_{+}=w1+\bar{w}e_{1234}+\psi^{\alpha}e_{\alpha}-\frac{1}{3!}(\star\bar{\psi})^{\alpha_1 \alpha_2 \alpha_3}e_{\alpha_1 \alpha_2 \alpha_3}+\chi^{p}e_{1p}-\frac{1}{2}\bar{\chi}^{r}\epsilon_r^{\,\,mn}\epsilon_{mn}\,,
\label{Majorana2}
\eea
where $\chi^{p}:=A^{1p}$, with $p=2,3,4$. Consider first the $su(3)$ transformation $S^{p\bar{q}}\Gamma_{p\bar{q}}$. Then
\bea
(S^{p\bar{q}}\Gamma_{p\bar{q}})e_{1r}=2 S^p{}_r e_{1p} \ .
\eea
Since the action of $SU(3)$ on $\mathbb{C}^3$ is simply transitive, without loss of generality we can impose $\chi^3=\chi^4=0$. Thus (\ref{Majorana2}) simplifies to
\bea
\phi_{+}^1=w1+\bar{w}e_{1234}+\psi^{\alpha}e_{\alpha}-\frac{1}{3!}(\star\bar{\psi})^{\alpha_1 \alpha_2 \alpha_3}e_{\alpha_1 \alpha_2 \alpha_3}+\chi e_{12}-\bar{\chi}e_{34}\,,
\label{phi3}
\eea
with $\chi:=\chi^1$. Next, let us consider the gauge transformations
\begin{eqnarray}
T_1:=\Gamma_{12}+\Gamma_{\bar{1}\bar{2}}\,,~~~T_{2}:=i(\Gamma_{12}-\Gamma_{\bar{1}\bar{2}}),~~~T_3:=\Gamma_{34}+\Gamma_{\bar{3}\bar{4}}\,,&T_{4}:=i(\Gamma_{34}-\Gamma_{\bar{3}\bar{4}})\,,
\label{Ts}
\end{eqnarray}
acting on
\begin{eqnarray}
v_1:=1+e_{1234}\,,&v_2:=i(1-e_{1234})\,,~~~v_{3}:=e_{12}-e_{34}\,,~~~v_{4}:=i(e_{12}+e_{34})\,.
\label{vectors}
\end{eqnarray}
It is easy to show that the representative matrices $M^{(a)}$ ($a=1,...,4$) of $T_{a}$ with respect to $v_{a}$ are elements of $so(4)$. In particular 
\bea
M^{(1)}-M^{(3)}=
\begin{pmatrix}
0 & 0 & 1 & 0 \\  0 & 0 & 0 & 0 \\ -1 & 0 & 0 & 0 \\ 0 & 0 & 0 & 0
\end{pmatrix}
\label{M1-M3}
\eea
and 
\bea
M^{(1)}+M^{(3)}=
\begin{pmatrix}
0 & 0 & 0 & 0 \\  0 & 0 & 0 & 1 \\ 0 & 0 & 0 & 0 \\ 0 & -1 & 0 & 0
\end{pmatrix}
\label{M1+M3}
\eea
are two copies of $so(2)$ acting on $\{v_1,v_3\}$ and $\{v_2,v_4\}$ respectively. The transitivity of $SO(2)$ on $\mathbb{R}^2$ allows us to set $\Re(\chi)=\Im(\chi)=0$, that is, $\chi=0$. Also, $i\Gamma_{1\bar{1}}$ is a $so(2)$ transformation acting on $\{v_1,v_2\}$. Thus, we can set $\Im(w)=0$, that is $w\in\mathbb{R}$. Combining the previous results, (\ref{phi3}) simplifies to
\bea
\phi_{+}^1=w(1+e_{1234})+\psi^{\alpha}e_{\alpha}-\frac{1}{3!}(\star\bar{\psi})^{\alpha_1 \alpha_2 \alpha_3}e_{\alpha_1 \alpha_2 \alpha_3}\,.
\label{phi4}
\eea
Next, consider a $su(4)$ transformation $\Lambda^{\alpha\bar{\beta}}\Gamma_{\alpha\bar{\beta}}$, where $\alpha=1,2,3,4$. We have that
\bea
(\Lambda^{\alpha\bar{\beta}}\Gamma_{\alpha\bar{\beta}})e_\lambda= 2 \Lambda^\alpha{}_\lambda e_\alpha \ .
\eea
Using the transitivity of $SU(4)$ on $\mathbb{C}^4$, we can set $\psi^2=\psi^3=\psi^4=0,  \,\,\psi:=\psi^1\in\mathbb{R}$. Thus (\ref{phi4}) 
boils down to
\bea
\phi_{+}^1=w(1+e_{1234})+\psi(e_{1}+e_{234})\,.
\label{10}
\eea
Moreover, $\Gamma_{\sharp 1}+\Gamma_{\sharp\bar{1}}$ is a $so(2)$ transformation on $1+e_{1234}$ and $e_{1}+e_{234}$, thus, using the transitivity 
of $SO(2)$, we can set $\psi=0$ in (\ref{10}), obtaining
\bea
\phi_{+}^1=w(1+e_{1234})\,.
\label{phi5}
\eea
Eventually, using $\parallel\phi_+^1\parallel^2=1$, we can set $w={1 \over \sqrt{2}}$ in (\ref{phi5}), obtaining (\ref{phi1}).

The stabilizer, or isotropy group, of $\phi_+^1$, with $\phi^1_+$ given by (\ref{phi1}), is $Spin(7)$. To see this, note that the stabilizer of $\phi_{+}^1$ is the subgroup of $Spin(9)$ whose generators satisfy
\bea
f^{ij}\Gamma_{ij}\,\phi_{+}^1=0\,.
\label{stabilizer1}
\eea
Using the oscillatory basis for the Gamma matrices
\bea
\Gamma_{\alpha}:=\frac{1}{\sqrt{2}}(\Gamma_{\alpha}-i\Gamma_{\alpha+5})\,,~~~\Gamma_{\bar{\alpha}}:=\frac{1}{\sqrt{2}}(\Gamma_{\alpha}+i\Gamma_{\alpha+5})\,,
\label{osc}
\eea
equation (\ref{stabilizer1}) boils down to
\bea
2f^{\alpha}_{\,\,\,\alpha}(e_{1234}-1)+\big(2f^{\alpha\beta}-\epsilon^{\alpha\beta\gamma\delta}f_{\gamma\delta}\big)e_{\alpha\beta}+2\sqrt{2}f^{\sharp\alpha}e_{\alpha}=0\,,
\eea
which, using the linear independence of the basis elements $e_{I}$, implies that
\bea
f_{\alpha}^{\,\,\,\alpha}=0\,\,\,\,,\,\,\,\,f_{\sharp\alpha}=0\,\,\,\,,\,\,\,\,\,f_{\bar{\alpha}\bar{\beta}}=\frac{1}{2}\epsilon_{\bar{\alpha}\bar{\beta}}^{\hspace{0.3cm}\,\,\gamma\delta}f_{\gamma\delta}\,, 
\label{spin7}
\eea
which is indeed the Lie algebra $spin(7)$ in a Hermitian basis \cite{fern}.

Using the oscillator basis (\ref{osc}), we have computed the $N=2$ linear system associated to the KSE (\ref{KSEN=2}), with $\phi_+^1$ given by (\ref{phi1}). The linear system is presented in Appendix A, and its solution is the following
\bea
\frac{1}{18}X_{\mu_1\mu_2\mu_3\mu_4}\epsilon^{\mu_1\mu_2\mu_3\mu_4}-\frac{1}{6}X_{\mu_1\mu_2}^{\,\,\,\,\,\,\,\,\,\,\,\mu_1\mu_2}=-\frac{1}{6}\Omega^{\mu}_{\,\,\,,\sharp\mu}-\frac{5}{6}\Omega_{\mu,\sharp}^{\,\,\,\,\,\,\,\mu}-\frac{2}{3}\Omega_{\sharp,\mu}^{\,\,\,\,\,\,\,\mu}\,,
\label{I}
\eea
\bea
X_{\alpha\bar{\beta}\mu}^{\,\,\,\,\,\,\,\,\,\,\,\,\mu}+\frac{1}{4}\delta_{\alpha\bar{\beta}}X_{\mu_1\mu_2}^{\,\,\,\,\,\,\,\,\,\,\,\mu_1\mu_2}=-2\Omega_{(\alpha,|\sharp|\bar{\beta})}+\frac{1}{4}\delta_{\alpha\bar{\beta}}\big(\Omega_{\mu,\sharp}^{\,\,\,\,\,\,\,\mu}+\Omega^{\mu}_{\,\,,\sharp\mu}\big)\,,
\label{II}
\eea
\bea
X_{(\alpha|\bar{\mu}_1\bar{\mu}_2\bar{\mu}_3}\epsilon^{\bar{\mu}_1\bar{\mu}_2\bar{\mu}_3}_{\,\,\,\,\,\,\,\,\,\,\,\,\,\,\,\,\,\,\,\,|\beta)}=6\Omega_{(\alpha,|\sharp|\beta)}\,,
\label{III}
\eea
\bea
X_{\alpha\beta\mu}^{\,\,\,\,\,\,\,\,\,\,\,\,\mu}+\frac{1}{2}X_{\bar{\mu}_1\bar{\mu}_2\nu}^{\,\,\,\,\,\,\,\,\,\,\,\,\,\,\,\,\,\nu}\epsilon^{\bar{\mu}_1\bar{\mu}_2}_{\,\,\,\,\,\,\,\,\,\,\,\alpha\beta}=2\Omega_{[\alpha,|\sharp|\beta]}+2\Omega_{\sharp,\alpha\beta}-\big(\Omega_{\sharp,\bar{\mu}_1\bar{\mu}_2}+\Omega_{\bar{\mu}_1,\sharp\bar{\mu}_2}\big)\epsilon^{\bar{\mu}_1\bar{\mu}_2}_{\,\,\,\,\,\,\,\,\,\,\,\,\alpha\beta}\,,
\label{IIItris}
\eea
\bea
X_{\sharp\alpha\beta\gamma}=-2\Omega_{[\alpha,\beta\gamma]}+\frac{2}{3}\big(-\Omega_{\nu,\,\,\,\bar{\mu}}^{\,\,\,\,\,\nu}+\Omega_{\bar{\mu},\nu}^{\,\,\,\,\,\,\,\,\,\,\nu}-\Omega_{\sharp,\sharp\bar{\mu}}\big)\epsilon^{\,\bar{\mu}}_{\,\,\,\alpha\beta\gamma}\,,
\label{IV}
\eea
\begin{eqnarray}
X_{\sharp\alpha\bar{\beta}\bar{\gamma}}&=&\frac{2}{3}(\Omega_{\alpha,\mu_1\mu_2}+\Omega_{\mu_1,\alpha\mu_2})\epsilon^{\mu_1\mu_2}_{\,\,\,\,\,\,\,\,\,\,\,\,\bar{\beta}\bar{\gamma}}-2\Omega_{\alpha,\bar{\beta}\bar{\gamma}}
\nonumber\\
&+&\frac{4}{3}\big(-\Omega_{\nu,\,\,\,\,[\bar{\beta}}^{\,\,\,\,\,\nu}+\Omega_{[\bar{\beta},|\nu|}^{\,\,\,\,\,\,\,\,\,\,\,\,\,\,\nu}+\frac{1}{2}\Omega_{\sharp,\sharp[\bar{\beta}}\big)\delta_{\bar{\gamma}]\alpha}\,,
\label{cinque}
\end{eqnarray}
\bea
Y_{\alpha\bar{\beta}}=2\Omega_{[\alpha,|\sharp|\bar{\beta}]}+\delta_{\alpha\bar{\beta}}\big(\Omega_{\sharp,\nu}^{\,\,\,\,\,\,\,\,\nu}+\frac{1}{2}\Omega^{\nu}_{\,\,\,,\sharp\nu}-\frac{1}{2}\Omega_{\nu,\sharp}^{\,\,\,\,\,\,\,\nu})\,,
\label{NEWNEW}
\eea
\bea
Y_{\sharp\alpha}=\frac{2}{3}\Omega_{\sharp,\sharp\alpha}+\frac{2}{3}\Omega_{\alpha,\nu}^{\,\,\,\,\,\,\,\,\nu}+\frac{2}{3}\Omega_{\bar{\mu}_1,\bar{\mu}_2\bar{\mu}_3}\epsilon^{\bar{\mu}_1\bar{\mu}_2\bar{\mu}_3}_{\,\,\,\,\,\,\,\,\,\,\,\,\,\,\,\,\,\,\,\alpha}-\frac{4}{3}\Omega^{\mu}_{\,\,\,,\mu\alpha}\,,
\label{VII}
\eea
\bea
Y_{\alpha\beta}=\big(\Omega_{\bar{\mu}_1,\sharp\bar{\mu}_2}-\Omega_{\sharp,\bar{\mu}_1\bar{\mu}_2}\big)\epsilon^{\bar{\mu}_1\bar{\mu}_2}_{\,\,\,\,\,\,\,\,\,\,\,\,\alpha\beta}+2\Omega_{\sharp,\alpha\beta}\,,
\label{VIII}
\eea
\bea
h_{\sharp}=-\frac{1}{2}(\Omega^{\mu}_{\,\,\,\,,\sharp\mu}+\Omega_{\mu,\sharp}^{\,\,\,\,\,\,\,\,\mu})\,,
\label{IX}
\eea
\bea
h_{\alpha}=\frac{1}{3}\Omega_{\sharp,\sharp\alpha}-\frac{2}{3}\Omega_{\alpha,\nu}^{\,\,\,\,\,\,\,\,\nu}-\frac{2}{3}\Omega_{\bar{\mu}_1,\bar{\mu}_2\bar{\mu}_3}\epsilon^{\bar{\mu}_1\bar{\mu}_2\bar{\mu}_3}_{\,\,\,\,\,\,\,\,\,\,\,\,\,\,\,\,\,\,\,\alpha}+\frac{4}{3}\Omega^{\mu}_{\,\,\,,\mu\alpha}\,.
\label{X}
\eea
After an extensive computation, the solution of the linear system, written in equations (\ref{I})-(\ref{X}) in a $SU(4)$ representation, can be repackaged in a compact $Spin(7)$ covariant form, using the $Spin(7)$ invariant differential forms
\bea
e^{\sharp}\,,~~~\tau:=\Re(\chi)-\frac{1}{2}\omega\wedge\omega\,,
\label{tauuu}
\eea
where $\tau$ is the $Spin(7)$ fundamental 4-form, $\chi$ and $\omega$ are 
a 4-form and a 2-form respectively, which in a local frame $e^{\alpha}$ read
\bea
\chi:=4e^{1234}\,,\,\,\,\,\,\,\,\,\omega:=-i\delta_{\alpha\bar{\beta}}e^{\alpha\bar{\beta}}\,.
\label{chiomega}
\eea
The $Spin(7)$ invariant forms (\ref{tauuu}) can be computed via the $Spin(7)$ gauge-invariant bilinears
\bea
\langle\phi_+^1,\Gamma_{i_1..i_n}\phi_+^1\rangle\,.
\eea
The $Spin(7)$ covariant solution of near-horizon geometries in $D=11$ supergravity preserving $N=2$ supersymmetries is given by 
\bea
h=-\frac{1}{3}\mathcal{L}_{e^{\sharp}}e^{\sharp}+\frac{1}{6}\theta^{(9)}_{\tau}-\frac{2}{3}(\star_9d\star_9 e^{\sharp})e^{\sharp}\,,
\label{finaleh}
\eea
\bea
Y=-d_he^{\sharp}-\frac{1}{48}(\mathcal{L}_{e^{\sharp}}\tau)_{i_1j_1j_2j_3}\tau^{j_1j_2j_3}_{\,\,\,\,\,\,\,\,\,\,\,\,\,\,\,\,i_2}e^{i_1i_2}\,,
\label{finaleY}
\eea
\begin{eqnarray}
X&=&-\frac{1}{7}(\star_9d\star_9 e^{\sharp})\tau-\frac{1}{4}\mathcal{L}_{e^{\sharp}}\tau+\frac{1}{4!}\big((de^{\sharp})_{i_1j}-\frac{3}{2}(\mathcal{L}_{e^{\sharp}}g)_{i_1j}\big)\tau^j_{\,\,\,i_2i_3i_4}e^{i_1i_2i_3i_4}
\nonumber \\
&-&e^{\sharp}\wedge\big\{\star_9d\star_9\tau+\frac{1}{3!}\big(-\frac{17}{24}(\mathcal{L}_{e^{\sharp}}e^{\sharp})_{j}+\frac{1}{6}(\theta^{(9)}_{\tau})_{j}\big)\tau^{j}_{\,\,\,\,i_1i_2i_3}e^{i_1i_2i_3}\big\}+X^{\textbf{27}}\,,
\nonumber \\
\label{XN=2}
\end{eqnarray}
where $\theta^{(9)}_{\tau}$ is the Lee form of $\tau$ on $\mathcal{S}$, i.e.
\bea
\theta^{(9)}_{\tau}:=\star_9(\tau\wedge\star_9d\tau)\,,
\label{Lee}
\eea
and $X^{\textbf{27}}\in\textbf{27}$ of $Spin(7)$ is projected out by the KSEs. The definition of the $\textbf{27}$ component of a 4-form on a 8-dimensional $Spin(7)$ manifold is given in (\ref{4form}). Furthermore, no geometric conditions arise.

\section{$N=4$ Solutions}

In this section we study near-horizon geometries preserving exactly $N=4$ supersymmetries, by obtaining the additional conditions on the geometry and the 4-form flux obtained by solving the 
KSE
\bea
\nabla_{i}^{(+)}\phi^2_{+}=0\,.
\label{KSEN=4}
\eea
The first step is to write the second (positive chirality) spinor $\phi_{+}^2$ in
the simplest possible canonical form. Using the residual $Spin(7)$ gauge invariance, which preserves the canonical
form of the first spinor $\phi^1_+$, the second spinor $\phi_{+}^2$ can be put in the following form
\bea
\phi_{+}^2=\frac{i}{\sqrt{2}}\cos\theta(1-e_{1234})+\frac{1}{\sqrt{2}}\sin\theta(e_{1}+e_{234})\,,
\label{phi+2}
\eea
where $\theta$ is a real function on $\cal{S}$. In order to show (\ref{phi+2}), we proceed similarly as for $\phi_+^1$. In particular, since $su(3)\subset spin(7)$, we can still simplify $\phi^2_+$ to
\bea
\phi_{+}^2=w1+\bar{w}e_{1234}+\psi^{\alpha}e_{\alpha}-\frac{1}{3!}(\star\bar{\psi})^{\alpha_1 \alpha_2 \alpha_3}e_{\alpha_1 \alpha_2 \alpha_3}+\chi e_{12}-\bar{\chi}e_{34}\,.
\label{phi3}
\eea
In order to simplify the spinor further, notice that  $M^{(1)}+M^{(3)}\in 
spin(7)$, while $M^{(1)}-M^{(3)}\notin spin(7)$ (the matrices $M^{(a)}$ are defined in the previous section). Thus we can only use $M^{(1)}+M^{(3)}$ to set $\Im(\chi)=0$. However
\bea
-M^{(2)}+M^{(4)}=
\begin{pmatrix}
0 & 0 & 0 & 0 \\  0 & 0 & 1 & 0 \\ 0 & -1 & 0 & 0 \\ 0 & 0 & 0 & 0
\end{pmatrix}
\label{-M2+M4}
\eea
is a $so(2)$ transformation acting on $\{v_2,v_3\}$, thus we can still set $\Re(\chi)=0$ and then $\chi=0$. However, differently from the first spinor, we cannot use $i\Gamma_{1\bar{1}}$ to set $w\in\mathbb{R}$ since $i\Gamma_{1\bar{1}}\notin spin(7)$, Thus
\bea
\phi_{+}^2=w1+\bar{w}e_{1234}+\psi^{\alpha}e_{\alpha}-\frac{1}{3!}(\star\bar{\psi})^{\alpha_1 \alpha_2 \alpha_3}e_{\alpha_1 \alpha_2 \alpha_3}\,,
\label{phi25}
\eea
Using the $su(4)\subset spin(7)$ gauge invariance, equation (\ref{phi25}) 
reduces to 
\bea
\phi_{+}^2=w_1\,(1+e_{1234})+iw_2(1-e_{1234})+\psi(e_{1}+e_{234})\,,
\label{phi22}
\eea
Eventually, using the orthonormality relation $\langle\phi_+^r,\phi_+^s\rangle=\delta^{rs}$, 
we can set $w_1=0$, $w_2=\frac{1}{\sqrt{2}}\cos\theta$ and $\psi=\frac{1}{\sqrt{2}}\sin\theta$ in (\ref{phi22}), with $\theta$ an arbitrary function of the horizon coordinates, obtaining (\ref{phi+2}).

The stabilizer of the two spinors $\phi_{+}^r$ ($r=1,2$), with $\phi_+^1$ given by (\ref{phi1}) and $\phi_+^2$ given by (\ref{phi+2}) is
\begin{itemize}
\item{$SU(3)$} if $\cos\theta,\sin\theta\ne 0$\,,
\item{$G_2$} if $\cos\theta=0$\,,
\item{$SU(4)$} if $\sin\theta=0$\,,
\end{itemize}
as we shall show now. The stabilizer of the two spinors is the subgroup of $Spin(7)$ whose generators satisfy
\bea
f^{ij}\Gamma_{ij}\phi_+^2=0\,,
\label{ss0}
\eea
with $f^{ij}\in spin(7)$. The direct computation shows that ($p,q,r=2,3,4)$
\begin{eqnarray}
f^{ij}\Gamma_{ij}\phi_+^2&=&\frac{i}{\sqrt{2}}\cos\theta\bigg(-2\epsilon^{pqr}(f_{r1}e_{pq}+f_{qr}e_{1p})\bigg)+\frac{1}{\sqrt{2}}\sin\theta\bigg(2\epsilon^{pqr}e_{1pq}(f_{1r}+f_{\bar{1}r})
\nonumber\\
&-&4e_p(f_1^{\,\,p}+f_{\bar{1}}^{\,\,p})+4f_p^{\,\,p}(e_1-e_{234})\bigg)=0\,,
\label{aux88}
\end{eqnarray}
where we used (\ref{spin7}). Let us assume now that both $\cos\theta$ and 
$\sin\theta$ are non-vanishing. Then, using the linear independence of the basis elements $e_I$, (\ref{aux88}) imply
\bea
f_{1p}=f_{1\bar{p}}=f_{pq}=f_{1\bar{1}}=f_p^{\,\,p}=0\,,
\eea
that is $f\in su(3)$. Now consider $\cos\theta=0$. Equations (\ref{spin7}) and (\ref{aux88}) imply
\begin{eqnarray}
f_{1\bar{1}}=f_p^{\,\,\,p}=0\,,~~~f_{1p}+f_{\bar{1}p}=0\,,~~~f_{1p}-f_{\bar{1}p}=\epsilon_p^{\,\,\bar{q}\bar{r}}f_{\bar{q}\bar{r}}\,,
\end{eqnarray}
that is $f\in \mathfrak{g}_2$, where $\mathfrak{g}_2$ is the Lie algebra of $G_2$. Eventually, let us consider $\sin\theta=0$. Equations (\ref{spin7}) and (\ref{aux88}) yield
\bea
f_{\alpha\beta}=0\,,\,\,\,\,\,f_{\alpha}^{\,\,\alpha}=0\,,
\eea
that is $f\in su(4)$.

In solving the KSE (\ref{KSEN=4}), the three cases outlined above must be treated separately.

\subsection{$SU(3)$ Isotropy Group}

In the $SU(3)$ case, the second spinor is given by (\ref{phi+2}), with both $\cos\theta$ and $\sin\theta$ non-vanishing.  The linear system associated to the second spinor is presented in Appendix B. After an extensive computation, the solution of the linear system can be expressed in terms of the $SU(3)$ gauge-invariant bilinears
\bea
\langle\phi_+^r,\Gamma_{i_1...i_n}\phi_+^s\rangle\,.
\label{bilifinale}
\eea
The non-trivial bilinears (\ref{bilifinale}) are the following
\bea
e^1\,,\,\,\,\,e^{\sharp}\,,\,\,\,\,\,\widehat{\omega}\,,\,\,\,\,\widehat{\chi}\,,
\eea
where $e^1$ is complex and
\bea
\widehat{\omega}:=-i\delta_{p\bar{q}}e^{p\bar{q}}\,,\,\,\,\,\,\,\widehat{\chi}:=2\sqrt{2}\,e^{234}\,
\label{hatomegahatchi}
\eea
are the usual $SU(3)$-invariant forms of 6-dimensional $SU(3)$-structures. Investigating the linear system (\ref{IS2})-(\ref{ULTS22}), it turns out that it is convenient to choose as 1-form bilinears the following orthonormal set
\begin{eqnarray}
K &:=& \frac{1}{\sqrt{2}}(e^1+e^{\bar{1}})
\nonumber \\
U&:=&\frac{i}{\sqrt{2}}\sin\theta(e^1-e^{\bar{1}})+\cos\theta\, e^{\sharp}
\nonumber \\
W&:=&\frac{i}{\sqrt{2}}\cos\theta(e^1-e^{\bar{1}})-\sin\theta\, e^{\sharp} \ ,
\label{Vthing}
\end{eqnarray}
rather than $e^1, e^{\bar{1}}$ and $e^{\sharp}$.

We have found the following (covariant) geometric conditions (hatted indices are real 6-dimensional $SU(3)$-indices)
\bea
(dK)_{\hat{i}\hat{j}}=0\,,
\label{dKij}
\eea
\bea
(dW)_{\hat{i}\hat{j}}=0\,,
\label{dWij}
\eea
\begin{eqnarray}
\sin\theta(dU)_{\hat{i}\hat{j}}-\frac{1}{2}\widehat{\omega}^{\hat{l}_1\hat{l}_2}\nabla_{[\hat{i}}\Re(\widehat{\chi})_{\hat{j}]\hat{l}_1\hat{l}_2}+\cos\theta\Re(\widehat{\chi})^{\hat{l}}_{\,\,\hat{i}\hat{j}}\big((\mathcal{L}_WK)_{\hat{l}}+[K,W]_{\hat{l}}\big)+\\+\Im(\widehat{\chi})^{\hat{l}}_{\,\,\hat{i}\hat{j}}\big[\frac{1}{2}(\theta^{(9)}_{\Re(\widehat{\chi})})_{\hat{l}}+\cos^2\theta\big((\mathcal{L}_UU)_{\hat{l}}+(\mathcal{L}_KK)_{\hat{l}}\big)-(\cos^2\theta+\frac{1}{2})(\theta^{(9)}_{\widehat{\omega}})_{\hat{l}}\big]=0\,,
\label{dVmessy1}
\end{eqnarray}
\bea
J_{(\hat{i}}^{\,\,\,\,\,\hat{l}}(\mathcal{L}_Ug)_{\hat{j})\hat{l}}+\frac{1}{2}\sin\theta J_{(\hat{i}}^{\,\,\,\,\,\hat{l}} M_{\hat{j})\hat{l}}=0\,,
\eea
\bea
(\mathcal{L}_UK)_{\hat{i}}=0\,,
\label{vecfin33}
\eea
\bea
(\mathcal{L}_UW)_{\hat{i}}=0\,,
\label{vecfin44}
\eea
\bea
(\mathcal{L}_WW)_{\hat{i}}-(\mathcal{L}_KK)_{\hat{i}}=0\,,
\label{vecfin5}
\eea
\bea
(\mathcal{L}_KW)_{\hat{i}}+(\mathcal{L}_WK)_{\hat{i}}=0\,,
\label{vecfin6}
\eea
\begin{eqnarray}
-\frac{1}{2}\sin\theta\,\widehat{\omega}^{\hat{j}_1\hat{j}_2}(\mathcal{L}_W\Re(\widehat{\chi}))_{\hat{j}_1\hat{j}_2\hat{i}}+\sin^2\theta(\mathcal{L}_WU)_{\hat{i}}+[W,U]_{\hat{i}}+\cos\theta J_{\hat{i}}^{\,\,\hat{j}}[K,U]_{\hat{j}}=0\,,
\label{vecfin7}
\end{eqnarray}
\begin{eqnarray}
-\frac{1}{2}\sin\theta\,\widehat{\omega}^{\hat{j}_1\hat{j}_2}(\mathcal{L}_K\Re(\widehat{\chi}))_{\hat{j}_1\hat{j}_2\hat{i}}+\sin^2\theta(\mathcal{L}_KU)_{\hat{i}}+[K,U]_{\hat{i}}-\cos\theta J_{\hat{i}}^{\,\,\hat{j}}[W,U]_{\hat{j}}=0\,,
\end{eqnarray}
\bea
\frac{1}{4}\,\widehat{\omega}^{\hat{j}_1\hat{j}_2}(\mathcal{L}_U\Re(\widehat{\chi}))_{\hat{j}_1\hat{j}_2\hat{i}}+\sin\theta\big(-(\mathcal{L}_UU)_{\hat{i}}+\frac{1}{2}(\theta^{(9)}_{\widehat{\omega}})_{\hat{i}}\big)=0\,,
\label{vecfin10}
\eea
\bea
2\cot\theta \partial_{\hat{i}}\theta-(\theta^{(9)}_{\Re(\widehat{\chi})})_{\hat{i}}+\frac{1}{2}\sin\theta (dU)_{\hat{j}_1\hat{j}_2}\Im(\widehat{\chi})^{\hat{j}_1\hat{j}_2}_{\,\,\,\,\,\,\,\,\,\,\,\hat{i}}=0\,,
\label{vecfin11}
\eea
\bea
\sin\theta\star_9(\Re(\widehat{\chi})\wedge\star_9d\widehat{\omega})-i_U\Theta=0\,,
\label{singfin1}
\eea
\bea
(1+\cos^2\theta)\star_9(\Im(\widehat{\chi})\wedge\star_9d\widehat{\omega})+\sin\theta i_U\Theta^{(9)}_{\Re(\widehat{\chi})}=0\,,
\label{singfin2}
\eea
\bea
i_U(\mathcal{L}_WW-\mathcal{L}_KK)=0\,,
\label{singfin6}
\eea
\bea
i_U(\mathcal{L}_{K}W+\mathcal{L}_{W}K)=0\,,
\label{singfin7}
\eea
\bea
(1+\cos^2\theta)\big(i_K\mathcal{L}_WK-\frac{1}{2}\star_9d\star_9W\big)+\frac{1}{4}\cos\theta\big(\cos\theta i_W\theta^{(9)}_{\Re(\widehat{\chi})}- i_K\Theta\big)=0\,,
\label{singfin8}
\eea
\bea
(1+\cos^2\theta)\big(i_W\mathcal{L}_KW-\frac{1}{2}\star_9d\star_9K\big)+\frac{1}{4}\cos\theta\big(\cos\theta i_K\theta^{(9)}_{\Re(\widehat{\chi})}+ i_W\Theta\big)=0\,,
\label{singfin9}
\eea
\bea
(1+\cos^2\theta)i_K\big(-\mathcal{L}_WW+\mathcal{L}_UU\big)+\frac{1}{2}\cos\theta i_W\Theta-\frac{1}{2}i_K\theta^{(9)}_{\Re(\widehat{\chi})}=0\,,
\label{singfin10}
\eea
\bea
(1+\cos^2\theta)\mathcal{L}_W\theta-\frac{1}{4}\sin(2\theta)i_W\theta^{(9)}_{\Re(\widehat{\chi})}+\frac{1}{2}\sin\theta i_K\Theta=0\,,
\label{singfin11}
\eea
\bea
(1+\cos^2\theta)\mathcal{L}_K\theta-\frac{1}{4}\sin(2\theta)i_K\theta^{(9)}_{\Re(\widehat{\chi})}-\frac{1}{2}\sin\theta i_W\Theta=0\,,
\label{singfin12}
\eea
\bea
(1+\cos^2\theta)\mathcal{L}_U\theta-\frac{1}{4}\sin(2\theta)i_U\theta^{(9)}_{\Re(\widehat{\chi})}=0\,,
\label{singfin13}
\eea
where
\bea
\theta^{(9)}_{\widehat{\omega}}:=\star_9(\widehat{\omega}\wedge\star_9d\widehat{\omega})\,,
\label{LEEE}
\eea
\bea
\theta^{(9)}_{\Re(\widehat{\chi})}:=\star_9(\Re(\widehat{\chi})\wedge\star_9d\Re(\widehat{\chi}))\,,
\label{LEEE}
\eea
are the Lee forms of $\widehat{\omega}$ and $\Re(\widehat{\chi})$ on $\mathcal{S}$ respectively, $J_{\hat{i}}^{\,\,\hat{j}}:= g^{\hat{j}\hat{l}}\widehat{\omega}_{\hat{i}\hat{l}}$ and
\bea
\Theta:=\star_9(\Re(\widehat{\chi})\wedge\star_9d\Im(\widehat{\chi}))\,,
\eea
\bea
M_{\hat{i}\hat{j}}:=-\widehat{\omega}^{l_1l_2}\nabla_{(\hat{i}}\Re(\widehat{\chi})_{\hat{j})l_1l_2}\,.  
\label{Mi1i2full}
\eea
Also, we have found a fully 9-dimensional covariant expression for $d\theta$, which expresses in a compact way the conditions (\ref{vecfin11}), (\ref{singfin11})-(\ref{singfin13}):
\begin{eqnarray}
(1+\cos^2\theta)\partial_i\theta&=&\frac{1}{4}\sin(2\theta)(\theta^{(9)}_{\Re(\widehat{\chi})})_{i}+\frac{1}{2}\tan\theta J_{i}^{\,\,\,j}\Theta_j
\nonumber\\
&-&\frac{1}{4}\frac{\sin^2\theta}{\cos\theta}(1+\cos^2\theta)(dU)_{j_1j_2}\Im(\widehat{\chi})^{j_1j_2}_{\,\,\,\,\,\,\,\,\,\,\,i}
\nonumber\\
&+&\frac{1}{2}\sin\theta\big((-i_K\Theta) W_i+(i_W\Theta) K_i\big) \ ,
\label{dtheta}
\end{eqnarray}
where $J_{i}^{\,\,j}\equiv g^{jl}\widehat{\omega}_{il}$. After having dealt with the geometric conditions, let us consider the fluxes. $h$ and $Y$ 
are already fixed by the $N=2$ solution, see (\ref{finaleh}) and (\ref{finaleY}), thus we have to analyse only $X$. It turns out that all of $X$ 
is fixed but the totally traceless (2,2) part, that is
\bea
X_{\bar{p}\bar{q}rs}^{(2,2)\textrm{TT}}:=X_{\bar{p}\bar{q}rs}-2X_{rt\,\,\,\,[\bar{q}}^{\,\,\,\,\,t}\delta_{\bar{p}]s}+2X_{st\,\,\,\,\,[\bar{q}}^{\,\,\,\,\,t}\delta_{\bar{p}]r}+X_{s_1s_2}^{\,\,\,\,\,\,\,\,\,\,s_1s_2}\delta_{r[\bar{p}}\delta_{\bar{q}]s}\,.
\label{TT}
\eea
The 4-form $X$ can be expressed in terms of the $SU(3)$-gauge invariant bilinears (\ref{bilifinale}) as follows
\begin{eqnarray}
X&=&K\wedge U\wedge W\wedge C_1+K\wedge U\wedge C_2+K\wedge W\wedge C_3+U\wedge W\wedge C_4+K\wedge C_5
\nonumber\\
&+&U\wedge C_6+W\wedge C_7+C_8\,,
\label{ansatz}
\end{eqnarray}
where $K$, $U$ and $W$ are defined in (\ref{Vthing}), $C_1\in\Omega^1(\mathcal{S})$, $C_2,C_3,C_4\in\Omega^2(\mathcal{S})$, $C_5,C_6,C_7\in\Omega^3(\mathcal{S})$ and $C_8\in\Omega^4(\mathcal{S})$ are fully expressed in terms of the bilinears (\ref{bilifinale}), with the exception of $C_8$, which  contains the unfixed $X^{(2,2)\textrm{TT}}$. The explicit expression of the $C_i$ ($i=1,2,..8$) is quite involved and we have preferred to 
list it in Appendix C.

To sum up, the full $SU(3)$-solution is given by
\begin{itemize}
\item{Geometric conditions:} equations  (\ref{dKij})-(\ref{singfin13})\,,
\item{$X$:} equations (\ref{ansatz})-(\ref{Y1full})\,,
\item{$Y$:} equation (\ref{finaleY})\,,
\item{$h$:} equation (\ref{finaleh})\,.
\end{itemize}
Equations (\ref{finaleY}) and (\ref{finaleh}) are written in $Spin(7)$ notation, but they can be broken down to $SU(3)$ straightforwardly; however, we did not perform explictly this computation, since it does not add any particular further insight.

\subsection{$G_2$ Isotropy Group}
In this section we solve the KSE (\ref{KSEN=4}) assuming $\cos\theta=0$ (thus $\sin\theta=1$), that is the second spinor (\ref{phi+2}) boils 
down to
\bea
\phi_{+}^2=\frac{1}{\sqrt{2}}(e_{1}+e_{234})\,.
\label{phi+G2}
\eea
The linear system associated to the second spinor is presented in Appendix B (set $\cos\theta=0$, $\sin\theta=1$ in equations (\ref{IS2})-(\ref{ULTS22})). The solution of the linear system, initially expressed in $SU(3)$ representations, after an extensive computation, can be written in terms of the $G_2$ gauge-invariant bilinears (\ref{bilifinale}), which are given by
\bea
e^{\sharp}\,,\,\,\,\,\,K\,,\,\,\,\phi\,,
\label{biliG2}
\eea
where $K$ is given by the first equation in (\ref{Vthing}) and $\phi$ is the fundamental $G_2$ 3-form, that is
\bea
\phi:=\Re(\widehat{\chi})+e^7\wedge\widehat{\omega}\,,
\label{G2fundamental}
\eea
where $\Re(\widehat{\chi})$ and $\widehat{\omega}$ are defined by (\ref{hatomegahatchi}) and $e^7\equiv\sqrt{2}\,\Im(e^1)$ is the seventh direction of the $G_2$ manifold. We can also construct the 4-form
\bea
\star_7\phi=-e^7\wedge\Im(\widehat{\chi})-\frac{1}{2}\widehat{\omega}\wedge\widehat{\omega}\,.
\label{psi}
\eea
We have found the following (covariant) geometric conditions ($\tilde{i},\tilde{j},...$ are 7-dimensional real $G_2$ indices)
\bea
(dK)_{\tilde{i}\tilde{j}}=0\,,
\label{G2finale1}
\eea
\bea
(de^{\sharp})_{\tilde{i}\tilde{j}}=0\,,
\label{G2finale2}
\eea
\bea
\phi^{\tilde{l}_1\tilde{l}_2\tilde{l}_3}\nabla_{(\tilde{i}}\star_7\phi_{\tilde{j})\tilde{l}_1\tilde{l}_2\tilde{l}_3}=0\,,
\label{G2finale3}
\eea
\bea
(\mathcal{L}_{e^{\sharp}}e^{\sharp})_{\tilde{i}}-(\mathcal{L}_KK)_{\tilde{i}}=0\,,
\label{projG23}
\eea
\bea
(\mathcal{L}_{K}e^{\sharp})_{\tilde{i}}+(\mathcal{L}_{e^{\sharp}}K)_{\tilde{i}}=0\,,
\label{projG24}
\eea
\bea
\phi^{\tilde{j}_1\tilde{j}_2\tilde{j}_3}(\mathcal{L}_{e^{\sharp}}\star_7\phi)_{\tilde{j}_1\tilde{j}_2\tilde{j}_3\tilde{i}}=0\,,
\eea
\bea
\phi^{\tilde{j}_1\tilde{j}_2\tilde{j}_3}(\mathcal{L}_{K}\star_7\phi)_{\tilde{j}_1\tilde{j}_2\tilde{j}_3\tilde{i}}=0\,,
\eea
\bea
(\theta^{(9)}_{\phi})_{\tilde{i}}=0\,,
\eea
\bea
[K,e^{\sharp}]_{\tilde{i}}-(\mathcal{L}_{K}e^{\sharp})_{\tilde{i}}=0\,,
\eea
\bea
(\star_7\phi)^{\tilde{j}_1\tilde{j}_2\tilde{j}_3\tilde{j}_4}(d\phi)_{\tilde{j}_1\tilde{j}_2\tilde{j}_3\tilde{j}_4}=0\,,
\eea
\bea
3i_K\mathcal{L}_{e^{\sharp}}e^{\sharp}+i_K\theta^{(9)}_{\phi}=0\,,
\label{projG21}
\eea
\bea
3i_{e^{\sharp}}\mathcal{L}_KK+i_{e^{\sharp}}\theta^{(9)}_{\phi}=0\,.
\label{projG22}
\eea
where 
\bea
\theta^{(9)}_{\phi}:=\star_9(\phi\wedge\star_9 d\phi)
\eea
is the Lee form of $\phi$ on $\mathcal{S}$.

Equations (\ref{G2finale1}), (\ref{G2finale2}), (\ref{projG23}), (\ref{projG24}), (\ref{projG21}) and (\ref{projG22}) imply that the exterior derivatives of $K$ and $e^{\sharp}$ on $\mathcal{S}$ can be written as follows
\bea
dK=\big(-\frac{1}{3} i_{e^{\sharp}}\theta^{(9)}_{\phi}\big)K\wedge e^{\sharp}+K\wedge\mathcal{L}_KK+e^{\sharp}\wedge\mathcal{L}_{e^{\sharp}}K\,,
\label{DKnice}
\eea
\bea
de^{\sharp}=\big(\frac{1}{3}i_K\theta^{(9)}_{\phi}\big)K\wedge e^{\sharp}-K\wedge\mathcal{L}_{e^{\sharp}}K+e^{\sharp}\wedge\mathcal{L}_KK\,.
\label{Desharpnice}
\eea
We can use (\ref{DKnice}) and (\ref{Desharpnice}) to introduce local coordinates on $\mathcal{S}$. In fact, equations (\ref{DKnice}) and (\ref{Desharpnice}) imply that $K$ and $e^{\sharp}$ form a 2-dimensional differential ideal on $\mathcal{S}$. Frobenius' theorem then ensures that locally we can choose a coordinate system \cite{extdif, frob}
\bea
y^{\tilde{i}},\,\,\,\,\,\,\,y^8:=v_1,\,\,\,\,\,\,\,y^9:=v_2
\eea
such that
\bea
K=g_1dv_1+g_2dv_2\,, \,\,\,\,\,\, e^{\sharp}=f_1dv_1+f_2dv_2\,,
\eea
for some functions $f_1,f_2,g_1$ and $g_2$. The slices $v_1=\textrm{const}$, $v_2=\textrm{const}$ are 2-dimensional integral manifolds, and every slice can be equipped with a $G_2$ structure with fundamental form given by
\bea
\phi=\frac{1}{3!}\phi_{l_1l_2l_3}(v_1,v_2,y^{\tilde{i}})dy^{l_1}\wedge dy^{l_2}\wedge dy^{l_3}\,.
\eea
It is worth noticing that the $(2+7)$-splitting of $\mathcal{S}$ which emerges in the $G_2$ case, there exists also in the $SU(3)$ case, although it is less transparent. In fact, the conditions (\ref{dKij}), (\ref{dWij}), (\ref{vecfin33}) and (\ref{vecfin44}) imply that $K$ and $W$ form a 2-dimensional ideal on $\mathcal{S}$. However, in the $SU(3)$ case we also have the 1-form bilinear $U$, but there are no simple conditions on $dU$, 
the only condition on it being (\ref{dVmessy1}). 

After having covariantized the geometric conditions, we are left to covariantize $X$. A somewhat long computation allows to express $X$ in terms of the $G_2$ gauge invariant bilinears (\ref{biliG2}) as follows
\begin{eqnarray}
X&=&K\wedge e^{\sharp}\wedge\bigg(\frac{1}{4!}\phi^{j_1j_2j_3}\nabla_{i_1}\star_7\phi_{i_2j_1j_2j_3}e^{i_1i_2}\bigg)+\frac{1}{4}\bigg(K\wedge(\mathcal{L}_{e^{\sharp}}g)_{i_1j}
\nonumber\\
&-&e^{\sharp}\wedge(\mathcal{L}_{K}g)_{i_1j}\bigg)\phi^j_{\,\,i_2i_3}e^{i_1i_2i_3}+X^{\textbf{27}}\,,
\label{XG2finale}
\end{eqnarray}
where $X^{\textbf{27}}\in\textbf{27}$ of $G_2$ is projected out by the KSEs. The definition of the $\textbf{27}$ component of a 4-form on a 7-dimensional $G_2$ manifold is given in (\ref{4formsG2}).
\subsection{$SU(4)$ Isotropy Group}
In this section we solve the KSE (\ref{KSEN=4}) assuming $\sin\theta=0$, i.e. the second spinor (\ref{phi+2}) boils down to
\bea
\phi_{+}^2=\frac{i}{\sqrt{2}}(1-e_{1234})\,.
\label{phisu4}
\eea
The linear system associated to the second spinor is given by
\bea
-\frac{1}{4}h_{\alpha}-\frac{1}{4}Y_{\sharp\alpha}+\frac{1}{2}\Omega_{\alpha,\mu}^{\,\,\,\,\,\,\,\,\mu}+\frac{1}{4}X_{\sharp\alpha\mu}^{\,\,\,\,\,\,\,\,\mu}=0\,,
\label{1finalesu4}
\eea
\bea
X_{\mu_1\mu_2\mu_3\mu_4}\epsilon^{\mu_1\mu_2\mu_3\mu_4}=0\,,
\label{2finalesu4}
\eea
\bea
\frac{1}{4}Y_{\alpha}^{\,\,\,\beta}-\frac{1}{2}\Omega_{\alpha,\sharp}^{\,\,\,\,\,\,\,\,\,\beta}-\frac{1}{4}X_{\alpha\mu}^{\,\,\,\,\,\,\mu\beta}+\frac{1}{12}\delta_{\alpha}^{\,\,\,\beta}\big(-Y_{\mu}^{\,\,\,\mu}-\frac{1}{2}X_{\mu_1\mu_2}^{\,\,\,\,\,\,\,\,\,\,\mu_1\mu_2}\big)=0\,,
\label{3finalesu4}
\eea
\bea
\frac{1}{2}\Omega_{\alpha,\beta_1\beta_2}+\frac{1}{12}X_{\sharp\alpha\beta_1\beta_2}=0\,,
\label{4finalesu4}
\eea
\begin{eqnarray}
\frac{1}{2}\Omega_{\alpha,\bar{\mu}_1\bar{\mu}_2}\epsilon^{\bar{\mu}_1\bar{\mu}_2}_{\,\,\,\,\,\,\,\,\,\,\beta_1\beta_2}+\frac{1}{4}X_{\sharp\alpha\bar{\mu}_1\bar{\mu}_2}\epsilon^{\bar{\mu}_1\bar{\mu}_2}_{\,\,\,\,\,\,\,\,\,\,\beta_1\beta_2}+\frac{1}{6}\epsilon_{\alpha\mu\beta_1\beta_2}\big(Y_{\sharp}^{\,\,\,\mu}-X_{\sharp\nu}^{\,\,\,\,\,\,\nu\mu}\big)=0\,,
\label{5finalesu4}
\end{eqnarray}
\bea
2\Omega_{\alpha,\beta\sharp}+\frac{1}{3}Y_{\alpha\beta}+\frac{1}{3}X_{\alpha\beta\mu}^{\,\,\,\,\,\,\,\,\,\,\mu}=0\,,
\label{6finalesu4}
\eea
\bea
-\frac{1}{3}Y_{\bar{\mu}_1\bar{\mu}_2}\epsilon^{\bar{\mu}_1\bar{\mu}_2}_{\,\,\,\,\,\,\,\,\,\,\,\alpha\beta}-\frac{1}{3}X_{\alpha\bar{\mu}_1\bar{\mu}_2\bar{\mu}_3}\epsilon^{\bar{\mu}_1\bar{\mu}_2\bar{\mu}_3}_{\,\,\,\,\,\,\,\,\,\,\,\,\,\,\,\,\,\beta}+\frac{1}{3}\epsilon_{\alpha\beta\mu_1\mu_2}X_{\nu}^{\,\,\,\nu\mu_1\mu_2}=0\,,
\label{7finalesu4}
\eea
\bea
-\frac{1}{2}h_{\alpha}-\frac{1}{6}Y_{\sharp\alpha}-\Omega_{\alpha,\mu}^{\,\,\,\,\,\,\,\,\mu}-\frac{1}{6}X_{\sharp\alpha\mu}^{\,\,\,\,\,\,\,\,\,\mu}=0\,,
\label{8finalesu4}
\eea
\bea
X_{\sharp\bar{\mu}_1\bar{\mu}_2\bar{\mu}_3}\epsilon^{\bar{\mu}_1\bar{\mu}_2\bar{\mu}_3}_{\,\,\,\,\,\,\,\,\,\,\,\,\,\,\,\,\,\,\,\,\alpha}=0\,,
\label{9finalesu4}
\eea
\bea
-\frac{1}{2}h_{\sharp}+\Omega_{\sharp,\mu}^{\,\,\,\,\,\,\,\mu}-\frac{1}{6}Y_{\mu}^{\,\,\,\mu}-\frac{1}{12}X_{\mu_1\mu_2}^{\,\,\,\,\,\,\,\,\,\,\,\,\mu_1\mu_2}=0\,,
\label{10finalesu4}
\eea
\bea
\frac{1}{3}Y_{\sharp\alpha}-\Omega_{\sharp,\sharp\alpha}+\frac{1}{3}X_{\sharp\alpha\mu}^{\,\,\,\,\,\,\,\,\mu}=0\,,
\label{11finalesu4}
\eea
\bea
\Omega_{\sharp,\alpha\beta}-\frac{1}{6}Y_{\alpha\beta}-\frac{1}{6}X_{\gamma\,\,\,\,\alpha\beta}^{\,\,\,\gamma}=0\,,
\label{12finalesu4}
\eea
\bea
\frac{1}{2}\Omega_{\sharp,\bar{\mu}_1\bar{\mu}_2}\epsilon^{\bar{\mu}_1\bar{\mu}_2}_{\,\,\,\,\,\,\,\,\,\,\,\,\alpha\beta}+\frac{1}{12}\epsilon^{\bar{\mu}_1\bar{\mu}_2}_{\,\,\,\,\,\,\,\,\,\,\,\alpha\beta}\big(-Y_{\bar{\mu}_1\bar{\mu}_2}+X_{\nu\,\,\,\,\bar{\mu}_1\bar{\mu}_2}^{\,\,\,\nu}\big)=0\,.
\label{13finalesu4}
\eea
Combining (\ref{1finalesu4})-(\ref{13finalesu4}) with the linear system of $\phi_+^1$, namely equation (\ref{IS})-(\ref{VIIS}), we can solve the linear system resulting from these two lots of equations and express the solution in terms of the $SU(4)$ gauge-invariant bilinears (\ref{bilifinale}), which are given by
\bea
e^{\sharp}\,,\,\,\,\,\,\omega\,,\,\,\,\,\chi\,,
\eea
where $\omega$ and $\chi$ are defined by (\ref{chiomega}). We have found the following geometric conditions ($\check{i},\check{j},\dots$ are 8-dimensional real $SU(4)$ indices)
\bea
(d\omega)^{\check{j}_1\check{j}_2\check{j}_3}\Re(\chi)_{\check{j}_1\check{j}_2\check{j}_3\check{i}}=0\,,
\label{cond1geo}
\eea
\bea
\frac{1}{2}(\star_9d\star_9\Re(\chi))_{\check{i}_1\check{i}_2\check{i}_3}+\big(\frac{1}{2}(\theta^{(9)}_{\omega})_{\check{j}}-(\mathcal{L}_{e^{\sharp}}e^{\sharp})_{\check{j}}\big)\Re(\chi)^{\check{j}}_{\,\,\,\,\check{i}_1\check{i}_2\check{i}_3}=0\,,
\label{cond2geo}
\eea
\bea
\frac{1}{12}\mathcal{L}_{e^{\sharp}}\Re(\chi)_{\check{j}_1\check{j}_2\check{j}_3[\check{i}_1}\Re(\chi)_{\check{i}_2]}^{\,\,\,\,\,\,\,\check{j}_1\check{j}_2\check{j}_3}+\omega_{\check{i}_1\check{i}_2}(de^{\sharp})_{\check{j}_1\check{j}_2}\omega^{\check{j}_1\check{j}_2}=0\,,
\label{cond3geo}
\eea
\bea
J_{(\check{i}_1}^{\,\,\,\,\,\,\check{j}}(\mathcal{L}_{e^{\sharp}}g)_{\check{i}_2)\check{j}}=0\,.
\eea
Moreover, the fluxes can be covariantized as follows
\bea
h=\frac{1}{2}(\star_9d\star_9 e^{\sharp})e^{\sharp}-\mathcal{L}_{e^{\sharp}}e^{\sharp}+\theta^{(9)}_{\omega}\,,
\eea
\bea
Y=-d_he^{\sharp}+\frac{1}{2}(de^{\sharp})_{j_1j_2}\omega^{j_1j_2}\,\omega\,,
\eea
\begin{eqnarray}
X&=&\frac{1}{2}e^{\sharp}\wedge\bigg\{\star_9d\star_9(\omega\wedge\omega)+\frac{1}{3!}\bigg((\theta^{(9)}_{\omega})_j-\frac{5}{4}(\mathcal{L}_{e^{\sharp}}e^{\sharp})_j\bigg)(\omega\wedge\omega)^j_{\,\,\,\,i_1i_2i_3}e^{i_1i_2i_3}\bigg\}
\nonumber\\
&-&S\wedge\omega+X^{\textrm{TT}(2,2)}\,,
\end{eqnarray}
where
\bea
\theta^{(9)}_{\Re(\chi)}:=\star_9(\Re(\chi)\wedge\star_9 d\Re(\chi))\,,
\eea
\bea
\theta^{(9)}_{\omega}:=\star_9(\omega\wedge\star_9 d\omega)\,,
\eea
are the Lee forms of $\Re(\chi)$ and $\omega$ on $\mathcal{S}$ respectively,
\bea
S_{ij}:= J_{[i}^{\,\,\,\,\,l}\big((de^{\sharp})_{j]l}+\frac{1}{2}(\mathcal{L}_{e^{\sharp}}g)_{j]l}\big)\,,
\eea
and $X^{\textrm{TT}(2,2)}$ is the $SU(4)$-traceless (2,2) part of $X$, that is
\bea
X^{\textrm{TT}(2,2)}_{\bar{\alpha}\bar{\beta}\gamma\delta}:= X_{\bar{\alpha}\bar{\beta}\gamma\delta}-X_{\gamma\lambda\,\,\,[\bar{\beta}}^{\,\,\,\,\,\lambda}\delta_{\bar{\alpha}]\delta}+X_{\delta\lambda\,\,\,[\bar{\beta}}^{\,\,\,\,\,\lambda}\delta_{\bar{\alpha}]\gamma}+\frac{1}{3}X_{\mu\nu}^{\,\,\,\,\,\,\,\mu\nu}\delta_{\gamma[\bar{\alpha}}\delta_{\bar{\beta}]\delta}\,.
\eea

\section{Integrability Conditions}

In this section we  shall show that all the components of the Einstein equation (\ref{Einstein}) are implied by the 11-dimensional KSE (\ref{KSE11dimensional}), the gauge field equation (\ref{Maxwell}) and the Bianchi identities $dF=0$. For a generic supersymmetric solution of $D=11$ supergravity, this is not always true. In particular, as shown in \cite{gpk},
for a $N=1$ supersymmetric solution which generates a null gauge-invariant isometry,
it is known that not all of the components of the Einstein equation are implied - one component
along the null-direction of the isometry must be imposed by hand.

However, for near-horizon solutions, we shall demonstrate that all the components of the Einstein equation are implied, irrespective of whether or 
not $\Delta$ vanishes. This analysis is purely local, and does not require any assumptions
on global properties of the solution. To investigate the integrability conditions, we assume that the KSE (\ref{KSE11dimensional}) are satisfied. Then the integrability conditions 
\bea
[\mathcal{D}_M,\mathcal{D}_N]\epsilon=0
\eea
yield \cite{Biran:1982eg}
\begin{eqnarray}
0&=&E_{MN}\Gamma^N\epsilon-\frac{1}{36}\star(d\star F+\frac{1}{2}F\wedge F)_{N_1N_2N_3}(\Gamma_{M}^{~~N_1N_2N_3}-6\delta_{M}^{~~N_1}\Gamma^{N_2N_3})\epsilon
\nonumber\\
&-&\frac{1}{6!}(dF)_{N_1N_2N_3N_4N_5}(\Gamma_M^{~~N_1N_2N_3N_4N_5}-10\delta_M^{~~N_1}\Gamma^{N_2N_3N_4N_5})\epsilon
\label{integrab}
\end{eqnarray}
where we have denoted 
\bea
E_{MN}:=R_{MN}-\frac{1}{12}F_{ML_1L_2L_3}F_{N}^{~~L_1L_2L_3}+\frac{1}{144}g_{MN}F^2\,.
\eea
Enforcing the gauge field equation (\ref{Maxwell}) and the Bianchi identities, (\ref{integrab}) boils down to
\bea
E_{MN}\Gamma^N\epsilon=0\,.
\eea

In the analysis of the integrability conditions, we shall not assume any of the results which
follow from the global assumptions made in the previous sections. In particular, on integrating
up the KSE along the lightcone directions, we find the following Killing spinors
\begin{eqnarray}
\epsilon_1= \phi_- + u \Gamma_+ \Theta_- \phi_- + ur \Gamma_- \Theta_+ \Gamma_+ \Theta_- \phi_-
\,,~~~~
\epsilon_2=\phi_++r\Gamma_-\Theta_+\phi_+\,,\,\,\,\,\,\,\,
\end{eqnarray}

Two separate cases must be considered. The first case corresponds to $\phi_-\ne 0$. Consider
\bea
E_{MN}\Gamma^N\epsilon_1=0\,,
\label{auz1}
\eea
where $M,N=+,-,i$. Setting $u=0$ in (\ref{auz1}), we obtain
\bea
E_{MN}\Gamma^N\phi_-=0\,,
\eea
which is equivalent to 
\bea
E_{M-}\Gamma_+\phi_-+E_{Mi}\Gamma^i\phi_-=0\,.
\label{auz2}
\eea
The first term in (\ref{auz2}) has positive (lightcone) chirality, while the second has negative (lightcone) chirality, thus (\ref{auz2}) holds iff both terms vanish independently, which in turn imply that
\bea
E_{M-}=E_{Mi}=0\,.
\eea
This means that all the components of $E_{MN}$ must vanish, apart from $E_{++}$. However, as is shown in \cite{Gutowski:2012eq}, in the near-horizon limit, $E_{++}=0$ is a consequence of the other bosonic field equations, thus indeed $E_{MN}=0$ for all $M,N$.

The second case corresponds to $\phi_-=0$ and $\phi_+\ne 0$. Consider
\bea
E_{MN}\Gamma^N\epsilon_2=0\,.
\eea
This equation boils down to 
\bea
E_{M+}\Gamma_-\phi_++2rE_{M-}\Theta_+\phi_++E_{Mi}\Gamma^i(\phi_++r\Gamma_-\Theta_+\phi_+)=0\,.
\label{auz3}
\eea
The first and the fourth terms in (\ref{auz3}) have negative (lightcone) chirality, while the second and the third ones have positive (lightcone) chirality, thus (\ref{auz2}) holds iff both terms vanish independently, which in turn imply that
\bea
E_{M+}\phi_+-rE_{Mi}\Gamma^i\Theta_+\phi_+=0
\label{auz4}
\eea
and
\bea
2rE_{M-}\Theta_+\phi_++E_{Mi}\Gamma^i\phi_+=0\,.
\label{auz5}
\eea
Taking the component $M=-$ of (\ref{auz4}), and exploiting the identities $E_{--}=0, E_{-i}=0$ (these expressions hold automatically for all 
near-horizon metrics), we get $E_{+-}=0$. Taking the component $M=j$ of (\ref{auz5}), we obtain $E_{ij}=0$. Taking the component $M=+$ of (\ref{auz5}), using $E_{+-}=0$, we get $E_{+i}=0$. Eventually, taking 
the $M=+$ component of (\ref{auz4}), we get $E_{++}=0$. Thus $E_{MN}=0$ for all $M,N$.

Thus, we have shown that for near-horizon geometries all the components of the Einstein equation (\ref{Einstein}) are implied by the 11-dimensional KSE (\ref{KSE11dimensional}), the gauge field equation (\ref{Maxwell}) and the Bianchi identities $dF=0$.

\section{Conclusion}

We have classified the conditions imposed on the geometry and 4-form flux
obtained from requiring that an extreme near-horizon geometry in $D=11$ 
supergravity
preserves $N=4$ supersymmetry. This analysis is tractable because previous global
analysis in \cite{mhor} has been used to reduce the calculation to that of solving
\begin{eqnarray}
\nabla_i^{(+)}\phi^r_+ =0, \qquad r=1,2
\end{eqnarray}
and for $N=4$ solutions, there is sufficient $Spin(9)$ gauge freedom present
to enable one to write $\{ \phi^1_+, \phi^2_+ \}$ in one of three simple canonical
forms, on exploiting the condition $\langle \phi^{r}_+, \phi^{s}_+ \rangle = \delta^{rs}$.
In particular, the common isotropy group of $\{ \phi^1_+, \phi^2_+ \}$ is 
one of three classes;
$SU(3)$, $G_2$ or $SU(4)$. For each class, the conditions on the geometry 
and fluxes have
been expressed in a fully gauge-invariant fashion, in terms of the gauge-invariant spinor bilinears  corresponding to each of these cases. 

There are a number of additional issues relevant to these $N=4$ solutions which would be interesting to explore further. First, although we have 
mentioned in section 2.3 that a generic 
$N=4$ near-horizon solution admits two commuting rotational isometries on ${\cal{S}}$, $\check{V}$ and $Z=Z^{12}$, it remains to determine how 
these isometries relate to the geometric structures
on ${\cal{S}}$ which we have derived in section 4. In practice, it is rather difficult to do this, because the isometries ${\check{V}}$ and $Z$ are constructed from both $\phi_+$ and $\phi_-$ spinors. We have used the majority of the gauge freedom to simplify the canonical forms of the $\phi_+$ spinors. In general, the $\phi_-$ spinors are rather complicated in form, as in the absence of useful gauge transformations to simplify these, 
one must instead utilize ({\ref{extragen}}) and ({\ref{implicit}}) to relate $\phi_+$ to $\phi_-$, and as a result there is a non-trivial appearance of various flux terms in the explicit expressions for the isometries.

Furthermore, it would be useful to see if there are any further conditions on the $N=4$ spinors from global analysis. In particular, we have seen that for the $SU(3)$ isotropy group case, there is a function $\theta$ which appears in the spinor $\phi^2_+$. Although the manner in which this 
function appears explicitly is dependent on the gauge choice we have made, one can straightforwardly write $\cos 2\theta$ in terms of gauge-invariant bilinears, and hence $\cos 2 \theta$ is a globally well-defined and smooth function on ${\cal{S}}$. In addition, the analysis of the KSE produces an expression for $d (\cos 2 \theta)$ in terms of various $SU(3)$ invariant bilinears ({\ref{dtheta}}). As we already have a constant norm condition,
$\langle \phi^{r}_+, \phi^{s}_+ \rangle = \delta^{rs}$ which is obtained via a (global) maximum principle argument, \cite{mhor}, it is natural to enquire if $\theta$ can be shown to be constant via an analogous analysis. However, on computing the Laplacian of $\cos 2 \theta$, there does not appear, a priori, to be any way of controlling the sign of the resulting terms in a way which is compatible with a maximum principle argument. We remark that there is also no immediate contradiction obtained if one assumes that $\theta$ is constant, with $\sin \theta \neq 0$, $\cos \theta \neq 0$; this would appear to be simply a special case of the more general analysis. It may also be the case that other properties of the geometry 
in the $N=4$ solutions could in principle be further constrained via similar global analysis, and it would be interesting to explore this further.

\setcounter{section}{0}
\setcounter{subsection}{0}

\appendix{$N=2$ $SU(4)$ KSE: linear system}
\label{N=2system}

The linear system obtained from the spinorial geometry analysis of the linear system 
obtained from the KSE of the $N=2$ supersymmetric near-horizon geometries is:

\bea
-\frac{1}{4}h_{\alpha}-\frac{1}{4}Y_{\sharp\alpha}+\frac{1}{2}\Omega_{\alpha,\mu}^{\,\,\,\,\,\,\,\,\,\,\mu}+\frac{1}{4}X_{\sharp\alpha\mu}^{\,\,\,\,\,\,\,\,\,\,\mu}=0\,,
\label{IS}
\eea
\begin{eqnarray}
&& Y_{\alpha}^{\,\,\,\beta}-2\Omega_{\alpha,\sharp}^{\,\,\,\,\,\,\,\,\,\beta}-X_{\alpha\mu}^{\,\,\,\,\,\,\,\,\mu\beta}+\frac{1}{3}\delta_{\alpha}^{\,\,\,\beta}\big(-Y_{\mu}^{\,\,\,\mu}-\frac{1}{2}X_{\mu_1\mu_2}^{\,\,\,\,\,\,\,\,\,\,\,\,\,\,\mu_1\mu_2}
\nonumber \\
&-&\frac{1}{12}X_{\mu_1\mu_2\mu_3\mu_4}\epsilon^{\mu_1\mu_2\mu_3\mu_4}\big)=0\,,
\end{eqnarray}
\begin{eqnarray}
&& \frac{1}{2}\Omega_{\alpha,\beta_1\beta_2}-\frac{1}{4}\Omega_{\alpha,\bar{\mu}_1\bar{\mu}_2}\epsilon^{\bar{\mu}_1\bar{\mu}_2}_{\,\,\,\,\,\,\,\,\,\,\,\,\beta_1\beta_2}-\frac{1}{8}X_{\sharp\alpha\bar{\mu}_1\bar{\mu}_2}\epsilon^{\bar{\mu}_1\bar{\mu}_2}_{\,\,\,\,\,\,\,\,\,\,\,\,\beta_1\beta_2}+\frac{1}{12}X_{\sharp\alpha\beta_1\beta_2}
\nonumber\\
&+&\frac{1}{12}\epsilon_{\alpha\mu\beta_1\beta_2}\big(-Y_{\sharp}^{\,\,\,\mu}+X_{\sharp\nu}^{\,\,\,\,\,\,\nu\mu}\big)=0\,,
\label{IIIS}
\end{eqnarray}
\bea
Y_{\alpha\beta}+\Omega_{\alpha,\beta\sharp}+Y_{\bar{\mu}_1\bar{\mu}_2}\epsilon^{\bar{\mu}_1\bar{\mu}_2}_{\,\,\,\,\,\,\,\,\,\,\,\,\,\alpha\beta}+X_{\alpha\bar{\mu}_1\bar{\mu}_2\bar{\mu}_3}\epsilon^{\bar{\mu}_1\bar{\mu}_2\bar{\mu}_3}_{\,\,\,\,\,\,\,\,\,\,\,\,\,\,\,\,\,\,\,\,\beta}+X_{\alpha\beta\mu}^{\,\,\,\,\,\,\,\,\,\,\,\,\mu}-\epsilon_{\alpha\beta\mu_1\mu_2}X_{\nu}^{\,\,\,\nu\mu_1\mu_2}=0\,,
\label{IVS}
\eea
\bea
-\frac{1}{4}h_{\alpha}-\frac{1}{12}Y_{\sharp\alpha}-\frac{1}{2}\Omega_{\alpha,\mu}^{\,\,\,\,\,\,\,\,\,\,\mu}-\frac{1}{12}X_{\sharp\alpha\mu}^{\,\,\,\,\,\,\,\,\,\,\,\mu}+\frac{1}{18}X_{\sharp\bar{\mu}_1\bar{\mu}_2\bar{\mu}_3}\epsilon^{\bar{\mu}_1\bar{\mu}_2\bar{\mu}_3}_{\,\,\,\,\,\,\,\,\,\,\,\,\,\,\,\,\,\,\,\,\alpha}=0\,,
\label{VS}
\eea
\bea
-\frac{1}{4}h_{\sharp}+\frac{1}{2}\Omega_{\sharp,\mu}^{\,\,\,\,\,\,\,\,\,\mu}-\frac{1}{12}Y_{\mu}^{\,\,\,\mu}-\frac{1}{24}X_{\mu_1\mu_2}^{\,\,\,\,\,\,\,\,\,\,\,\,\mu_1\mu_2}+\frac{1}{72}X_{\mu_1\mu_2\mu_3\mu_4}\epsilon^{\mu_1\mu_2\mu_3\mu_4}=0\,,
\label{VIS}
\eea
\bea
\frac{1}{6}Y_{\sharp\alpha}-\frac{1}{2}\Omega_{\sharp,\sharp\alpha}+\frac{1}{6}X_{\sharp\alpha\mu}^{\,\,\,\,\,\,\,\,\,\,\mu}+\frac{1}{18}X_{\sharp\bar{\mu}_1\bar{\mu}_2\bar{\mu}_3}\epsilon^{\bar{\mu}_1\bar{\mu}_2\bar{\mu}_3}_{\,\,\,\,\,\,\,\,\,\,\,\,\,\,\,\,\,\,\alpha}=0\,,
\label{VIIS}
\eea
\bea
\frac{1}{2}\Omega_{\sharp,\alpha\beta}-\frac{1}{4}\Omega_{\sharp,\bar{\mu}_1\bar{\mu}_2}\epsilon^{\bar{\mu}_1\bar{\mu}_2}_{\,\,\,\,\,\,\,\,\,\,\,\,\alpha\beta}-\frac{1}{12}Y_{\alpha\beta}-\frac{1}{12}X_{\gamma\,\,\,\,\alpha\beta}^{\,\,\,\gamma}+\frac{1}{24}\epsilon^{\bar{\mu}_1\bar{\mu}_2}_{\,\,\,\,\,\,\,\,\,\,\,\alpha\beta}\big(Y_{\bar{\mu}_1\bar{\mu}_2}-X_{\nu\,\,\,\,\bar{\mu}_1\bar{\mu}_2}^{\,\,\,\nu}\big)=0\,. \ \
\label{VIIIS}
\eea

\appendix{$N=4$ $SU(3)$ KSE: linear system}
\label{SU3}
The $N=4$ linear system associated to the KSE (\ref{KSEN=4}), with $\phi_+^2$ given by (\ref{phi+2}), for both $\cos\theta$ and $\sin\theta$ non-vanishing, is given by ($p=2,3,4$)
\begin{eqnarray}
&& \frac{i}{\sqrt{2}}\partial_1\cos\theta+\frac{i}{\sqrt{2}}\cos\theta\bigg(-\frac{1}{4}h_1-\frac{1}{4}Y_{\sharp1}+\frac{1}{2}\Omega_{1,1\bar{1}}+\frac{1}{2}\Omega_{1,p}^{\,\,\,\,\,\,\,p}+\frac{1}{4}X_{\sharp1p}^{\,\,\,\,\,\,\,\,p}\bigg)
\nonumber\\
&+&\sin\theta\bigg(\frac{1}{2}\Omega_{1,\sharp1}+\frac{1}{12}X_{1pqr}\epsilon^{pqr}\bigg)=0\,,
\label{IS2}
\end{eqnarray}
\begin{eqnarray}
&& i\cos\theta\bigg(\frac{1}{6}Y_{1\bar{1}}-\frac{1}{2}\Omega_{1,\sharp\bar{1}}-\frac{1}{12}Y_p^{\,\,\,p}-\frac{1}{6}X_{1\bar{1}p}^{\,\,\,\,\,\,\,\,\,p}-\frac{1}{24}X_{pq}^{\,\,\,\,\,pq}+\frac{1}{36}X_{1pqr}\epsilon^{pqr}\bigg)
\nonumber\\
&+&\frac{1}{\sqrt{2}}\partial_1\sin\theta+\frac{1}{\sqrt{2}}\sin\theta\bigg(-\frac{1}{4}h_1+\frac{1}{12}Y_{\sharp1}+\frac{1}{2}\Omega_{1,p}^{\,\,\,\,\,\,\,p}
\nonumber\\
&-&\frac{1}{2}\Omega_{1,1\bar{1}}-\frac{1}{12}X_{\sharp1p}^{\,\,\,\,\,\,\,\,\,p}-\frac{1}{18}X_{\sharp pqr}\epsilon^{pqr}\bigg)=0\,,
\label{IIS2}
\end{eqnarray}
\begin{eqnarray}
&& i\cos\theta\bigg(\frac{1}{4}Y_1^{\,\,p}-\frac{1}{2}\Omega_{1,\sharp}^{\,\,\,\,\,\,\,\,p}-\frac{1}{4}X_{1t}^{\,\,\,\,\,tp}\bigg)+\frac{1}{\sqrt{2}}\sin\theta\bigg(-\Omega_{1,1}^{\,\,\,\,\,\,\,p}
\nonumber\\
&-&\frac{1}{2}\Omega_{1,qr}\epsilon^{pqr}+\frac{1}{4}X_{\sharp 1qr}\epsilon^{pqr}\bigg)=0\,,
\label{IIIS2}
\end{eqnarray}
\begin{eqnarray}
&& \frac{i}{\sqrt{2}}\cos\theta\bigg(\Omega_{1,\bar{1}}^{\,\,\,\,\,\,\,\,p}+\frac{1}{6}Y_{\sharp}^{\,\,\,p}+\frac{1}{3}X_{\sharp1\bar{1}}^{\,\,\,\,\,\,\,\,p}-\frac{1}{6}X_{\sharp q}^{\,\,\,\,\,qp}+\frac{1}{2}\Omega_{1,qr}\epsilon^{pqr}
\nonumber\\
&+&\frac{1}{12}X_{\sharp 1qr}\epsilon^{pqr}\bigg)+\sin\theta\bigg(-\frac{1}{12}Y_{1}^{\,\,p}-\frac{1}{2}\Omega_{1,\sharp}^{\,\,\,\,\,\,\,\,p}+\frac{1}{12}Y_{qr}\epsilon^{pqr}
\nonumber\\
&+&\frac{1}{12}X_{1t}^{\,\,\,\,\,tp}+\frac{1}{6}X_{1\bar{1}qr}\epsilon^{pqr}+\frac{1}{36}X_{qrs}^{\,\,\,\,\,\,\,\,\,p}\epsilon^{qrs}\bigg)=0\,,
\label{IVS2}
\end{eqnarray}
\begin{eqnarray}
&& \frac{i}{\sqrt{2}}\cos\theta\bigg(\frac{1}{2}\Omega_{1,}^{\,\,\,\,qr}\epsilon_{pqr}+\frac{1}{4}X_{\sharp1}^{\,\,\,\,\,\,qr}\epsilon_{pqr}+\Omega_{1,1p}\bigg)
\nonumber\\
&+&\sin\theta\bigg(\frac{1}{4}Y_{1p}+\frac{1}{2}\Omega_{1,\sharp p}+\frac{1}{4}X_{1pq}^{\,\,\,\,\,\,\,\,\,q}\bigg)=0\,,
\label{VS2}
\end{eqnarray}
\begin{eqnarray}
&& i\cos\theta\bigg(-\frac{1}{6}X_{1\bar{1}}^{\,\,\,\,\,\,qr}\epsilon_{pqr}-\frac{1}{12}Y^{qr}\epsilon_{pqr}+\frac{1}{12}X_{s}^{\,\,\,sqr}\epsilon_{pqr}+\frac{1}{12}Y_{1p}
\nonumber\\
&-&\frac{1}{2}\Omega_{1,\sharp p}+\frac{1}{12}X_{1pq}^{\,\,\,\,\,\,\,\,\,q}\bigg)+\frac{1}{\sqrt{2}}\sin\theta\bigg(\frac{1}{2}\Omega_{1,}^{\,\,\,\,qr}\epsilon_{pqr}+\Omega_{1,\bar{1}p}-\frac{1}{6}Y_{\sharp p}
\nonumber\\
&-&\frac{1}{12}X_{\sharp 1}^{\,\,\,\,\,qr}\epsilon_{pqr}-\frac{1}{3}X_{\sharp p1\bar{1}}-\frac{1}{6}X_{\sharp pq}^{\,\,\,\,\,\,\,q}\bigg)=0\,.
\label{VIS2}
\end{eqnarray}
\begin{eqnarray}
&& i\cos\theta\bigg(-\frac{1}{12}X_1^{\,\,pqr}\epsilon_{pqr}+\frac{1}{2}\Omega_{1,\sharp 1}\bigg)+\frac{1}{\sqrt{2}}\partial_1\sin\theta
\nonumber\\
&+&\frac{1}{\sqrt{2}}\sin\theta\bigg(-\frac{1}{4}h_1+\frac{1}{4}Y_{\sharp 
1}+\frac{1}{2}\Omega_{1,1\bar{1}}-\frac{1}{2}\Omega_{1,p}^{\,\,\,\,\,\,\,p}+\frac{1}{4}X_{\sharp1p}^{\,\,\,\,\,\,\,\,p}\bigg)=0\,,
\label{VIIS2}
\end{eqnarray}
\begin{eqnarray}
&& -\frac{i}{\sqrt{2}}\partial_1\cos\theta+\frac{i}{\sqrt{2}}\cos\theta\bigg(-\frac{1}{18}X_{\sharp}^{\,\,\,pqr}\epsilon_{pqr}+\frac{1}{4}h_1+\frac{1}{12}Y_{\sharp 1}+\frac{1}{2}\Omega_{1,1\bar{1}}
\nonumber\\
&+&\frac{1}{2}\Omega_{1,p}^{\,\,\,\,\,\,\,p}+\frac{1}{12}X_{\sharp1p}^{\,\,\,\,\,\,\,\,p}\bigg)+\sin\theta\bigg(\frac{1}{6}Y_{1\bar{1}}+\frac{1}{2}\Omega_{1,\sharp\bar{1}}+\frac{1}{12}Y_p^{\,\,p}
\nonumber\\
&+&\frac{1}{6}X_{1\bar{1}p}^{\,\,\,\,\,\,\,\,p}+\frac{1}{36}X_{1}^{\,\,pqr}\epsilon_{pqr}-\frac{1}{24}X_{pq}^{\,\,\,\,\,pq}\bigg)=0\,,
\label{VIIIS2}
\end{eqnarray}
\begin{eqnarray}
&& \frac{i}{\sqrt{2}}\partial_p\cos\theta+\frac{i}{\sqrt{2}}\cos\theta\bigg(-\frac{1}{4}h_p-\frac{1}{4}Y_{\sharp p}+\frac{1}{2}\Omega_{p,1\bar{1}}+\frac{1}{2}\Omega_{p,q}^{\,\,\,\,\,\,\,q}
\nonumber\\
&+&\frac{1}{4}X_{\sharp p1\bar{1}}+\frac{1}{4}X_{\sharp pq}^{\,\,\,\,\,\,\,q}\bigg)+\sin\theta\bigg(\frac{1}{4}Y_{p1}+\frac{1}{2}\Omega_{p,\sharp 1}+\frac{1}{4}X_{1pq}^{\,\,\,\,\,\,\,\,\,q}\bigg)=0\,,
\label{IXS2}
\end{eqnarray}
\label{IXS2}
\begin{eqnarray}
&& i\cos\theta\bigg(\frac{1}{4}Y_{p\bar{1}}-\frac{1}{2}\Omega_{p,\sharp\bar{1}}+\frac{1}{4}X_{\bar{1}pq}^{\,\,\,\,\,\,\,\,\,q}\bigg)+\frac{1}{\sqrt{2}}\partial_p\sin\theta
\nonumber\\
&+&\frac{1}{\sqrt{2}}\sin\theta\bigg(-\frac{1}{4}h_p+\frac{1}{4}Y_{\sharp 
p}+\frac{1}{2}\Omega_{p,q}^{\,\,\,\,\,\,\,q}-\frac{1}{2}\Omega_{p,1\bar{1}}+\frac{1}{4}X_{\sharp p1\bar{1}}-\frac{1}{4}X_{\sharp pq}^{\,\,\,\,\,\,\,\,q}\bigg)=0\, ,
\nonumber \\
\label{XS2}
\end{eqnarray}
\begin{eqnarray}
&& i\cos\theta\bigg(\frac{1}{4}Y_p^{\,\,t}-\frac{1}{2}\Omega_{p,\sharp}^{\,\,\,\,\,\,\,t}-\frac{1}{12}\delta_p^{\,\,\,t}Y_{1\bar{1}}-\frac{1}{12}\delta_p^{\,\,\,t}Y_s^{\,\,s}-\frac{1}{4}X_{1\bar{1}p}^{\,\,\,\,\,\,\,\,\,\,t}
\nonumber\\
&+&\frac{1}{4}X_{pq}^{\,\,\,\,\,tq}+\frac{1}{12}\delta_p^{\,\,\,t}X_{1\bar{1}q}^{\,\,\,\,\,\,\,\,\,q}-\frac{1}{24}\delta_p^{\,\,\,t}X_{mn}^{\,\,\,\,\,\,\,\,mn}+\frac{1}{36}\delta_p^{\,\,\,t}X_{1qrs}\epsilon^{qrs}\bigg)
\nonumber\\
&+&\frac{1}{\sqrt{2}}\sin\theta\bigg(-\Omega_{p,1}^{\,\,\,\,\,\,\,\,t}-\frac{1}{2}\Omega_{p,qr}\epsilon^{qrt}-\frac{1}{6}\delta_p^{\,\,\,t}Y_{\sharp 1}-\frac{1}{2}X_{\sharp 1 p }^{\,\,\,\,\,\,\,\,\,t}+\frac{1}{4}X_{\sharp pqr}\epsilon^{qrt}
\nonumber\\
&+&\frac{1}{6}\delta_{p}^{\,\,\,t}X_{\sharp 1q}^{\,\,\,\,\,\,\,q}-\frac{1}{18}\delta_{p}^{\,\,\,t}X_{\sharp qrs}\epsilon^{qrs}\bigg)=0\,,
\label{SING1}
\end{eqnarray}
\begin{eqnarray}
&& \frac{i}{\sqrt{2}}\cos\theta\bigg(\Omega_{p,\bar{1}}^{\,\,\,\,\,\,\,\,\,t}-\frac{1}{6}\delta_{p}^{\,\,\,t}Y_{\sharp\bar{1}}-\frac{1}{2}X_{\sharp\bar{1} p}^{\,\,\,\,\,\,\,\,\,t}+\frac{1}{6}\delta_p^{\,\,t}X_{\sharp\bar{1}q}^{\,\,\,\,\,\,\,q}+\frac{1}{2}\Omega_{p,qr}\epsilon^{qrt}
\nonumber\\
&+&\frac{1}{4}X_{\sharp pqr}\epsilon^{qrt}-\frac{1}{18}\delta_p^{\,\,t}X_{\sharp qrs}\epsilon^{qrs}\bigg)+\sin\theta\bigg(-\frac{1}{4}Y_p^{\,\,t}-\frac{1}{2}\Omega_{p,\sharp}^{\,\,\,\,\,\,\,\,t}
\nonumber\\
&+&\frac{1}{12}\delta_{p}^{\,\,\,t}Y_s^{\,\,s}-\frac{1}{12}\delta_{p}^{\,\,\,t}Y_{1\bar{1}}-\frac{1}{4}X_{1\bar{1}p}^{\,\,\,\,\,\,\,\,\,t}-\frac{1}{4}X_{pq}^{\,\,\,\,\,tq}-\frac{1}{4}X_{\bar{1}pqr}\epsilon^{qrt}
\nonumber\\
&+&\frac{1}{12}\delta_{p}^{\,\,\,t}X_{1\bar{1}q}^{\,\,\,\,\,\,\,\,\,q}+\frac{1}{24}\delta_{p}^{\,\,\,t}X_{mn}^{\,\,\,\,\,\,\,\,mn}+\frac{1}{18}\delta_{p}^{\,\,\,t}X_{\bar{1}pqr}\epsilon^{pqr}\bigg)=0\,,
\label{SING2}
\end{eqnarray}
\begin{eqnarray}
&& \frac{i}{\sqrt{2}}\cos\theta\bigg(\frac{1}{2}\Omega_{p,}^{\,\,\,\,\,qr}\epsilon_{qrs}+\frac{1}{4}X_{\sharp p}^{\,\,\,\,\,\,qr}\epsilon_{qrs}+\frac{1}{6}\epsilon_{prs}Y_{\sharp}^{\,\,\,r}-\frac{1}{6}\epsilon_{prs}X_{\sharp1\bar{1}}^{\,\,\,\,\,\,\,\,\,r}-\frac{1}{6}\epsilon_{prs}X_{\sharp t}^{\,\,\,\,\,\,tr}
\nonumber\\
&-&\Omega_{p,s1}-\frac{1}{6}X_{\sharp1ps}\big)+\sin\theta\big(\frac{1}{12}Y_{ps}+\frac{1}{2}\Omega_{p,\sharp s}+\frac{1}{6}\epsilon_{prs}Y_1^{\,\,\,r}-\frac{1}{4}X_{p1}^{\,\,\,\,\,qr}\epsilon_{qrs}
\nonumber\\
&-&\frac{1}{12}X_{1\bar{1}ps}+\frac{1}{12}X_{psr}^{\,\,\,\,\,\,\,\,r}-\frac{1}{6}\epsilon_{prs}X_{1t}^{\,\,\,\,\,\,tr}\bigg)=0\,,
\label{2T}
\end{eqnarray}
\begin{eqnarray}
&& i\cos\theta\bigg(-\frac{1}{4}X_{p\bar{1}}^{\,\,\,\,\,qr}\epsilon_{qrs}+\frac{1}{6}\epsilon_{prs}Y_{\bar{1}}^{\,\,\,r}-\frac{1}{6}\epsilon_{prs}X_{\bar{1}t}^{\,\,\,\,\,tr}+\frac{1}{12}Y_{ps}-\frac{1}{2}\Omega_{p,\sharp s}
\nonumber\\
&+&\frac{1}{12}X_{psr}^{\,\,\,\,\,\,\,\,r}+\frac{1}{12}X_{1\bar{1}ps}\bigg)+\frac{1}{\sqrt{2}}\sin\theta\bigg(\frac{1}{2}\Omega_{p,}^{\,\,\,\,\,qr}\epsilon_{qrs}-\Omega_{p,s\bar{1}}-\frac{1}{6}Y_{\sharp}^{\,\,\,r}\epsilon_{prs}
\nonumber\\
&-&\frac{1}{4}X_{\sharp p}^{\,\,\,\,\,qr}\epsilon_{qrs}+\frac{1}{6}X_{\sharp\bar{1} ps}-\frac{1}{6}X_{\sharp 1\bar{1}}^{\,\,\,\,\,\,\,q}\epsilon_{pqs}+\frac{1}{6}\epsilon_{pqs}X_{\sharp r}^{\,\,\,\,\,\,rq}\bigg)=0\,,
\label{2Tbis}
\end{eqnarray}
\begin{eqnarray}
&& i\cos\theta\bigg(-\frac{1}{12}Y^{qr}\epsilon_{pqr}-\frac{1}{12}X_p^{\,\,qrs}\epsilon_{qrs}+\frac{1}{12}X_{1\bar{1}}^{\,\,\,\,\,qr}\epsilon_{pqr}+\frac{1}{12}X_s^{\,\,sqr}\epsilon_{pqr}
\nonumber\\
&+&\frac{1}{12}Y_{1p}+\frac{1}{2}\Omega_{p,\sharp 1}+\frac{1}{12}X_{1pq}^{\,\,\,\,\,\,\,\,\,q}\bigg)+\frac{1}{\sqrt{2}}\partial_{p}\sin\theta+\frac{1}{\sqrt{2}}\sin\theta\big(-\frac{1}{4}h_p
\nonumber\\
&+&\frac{1}{12}Y_{\sharp p}+\frac{1}{2}\Omega_{p,1\bar{1}}-\frac{1}{2}\Omega_{p,q}^{\,\,\,\,\,\,\,q}-\frac{1}{12}X_{\sharp p1\bar{1}}+\frac{1}{12}X_{\sharp pq}^{\,\,\,\,\,\,\,\,\,q}+\frac{1}{6}X_{\sharp 1}^{\,\,\,\,\,\,qr}\epsilon_{pqr}\bigg)=0\,,
\label{WW}
\end{eqnarray}
\begin{eqnarray}
&& -\frac{i}{\sqrt{2}}\partial_p\cos\theta+\frac{i}{\sqrt{2}}\cos\theta\bigg(\frac{1}{6}X_{\sharp\bar{1}}^{\,\,\,\,\,\,qr}\epsilon_{pqr}+\frac{1}{4}h_p+\frac{1}{12}Y_{\sharp p}+\frac{1}{2}\Omega_{p,1\bar{1}}
\nonumber\\
&+&\frac{1}{2}\Omega_{p,q}^{\,\,\,\,\,\,\,q}+\frac{1}{12}X_{\sharp p1\bar{1}}+\frac{1}{12}X_{\sharp pq}^{\,\,\,\,\,\,\,\,q}\bigg)+\sin\theta\bigg(-\frac{1}{12}Y_{\bar{1}p}+\frac{1}{2}\Omega_{p,\sharp\bar{1}}+\frac{1}{12}Y^{qr}\epsilon_{pqr}
\nonumber\\
&-&\frac{1}{12}X_{\bar{1}pq}^{\,\,\,\,\,\,\,q}+\frac{1}{12}X_p^{\,\,qrs}\epsilon_{qrs}+\frac{1}{12}X_{1\bar{1}}^{\,\,\,\,\,qr}\epsilon_{pqr}-\frac{1}{12}X_t^{\,\,tqr}\epsilon_{pqr}\bigg)=0\,,
\label{fullsystemp}
\end{eqnarray}
\begin{eqnarray}
&& \frac{i}{\sqrt{2}}\partial_{\sharp}\cos\theta+\frac{i}{\sqrt{2}}\cos\theta\bigg(-\frac{1}{4}h_{\sharp}+\frac{1}{2}\Omega_{\sharp,1\bar{1}}+\frac{1}{2}\Omega_{\sharp, p}^{\,\,\,\,\,\,\,\,p}-\frac{1}{12}Y_{1\bar{1}}
\nonumber\\
&-&\frac{1}{12}Y_p^{\,\,p}+\frac{1}{12}X_{1\bar{1}p}^{\,\,\,\,\,\,\,\,\,p}-\frac{1}{24}X_{pq}^{\,\,\,\,\,pq}-\frac{1}{18}X_{1pqr}\epsilon^{pqr}\bigg)
\nonumber\\
&+&\sin\theta\bigg(\frac{1}{6}Y_{\sharp1}+\frac{1}{2}\Omega_{\sharp,\sharp1}-\frac{1}{6}X_{\sharp1p}^{\,\,\,\,\,\,\,\,p}+\frac{1}{18}X_{\sharp pqr}\epsilon^{pqr}\bigg)=0\,,
\label{SING3}
\end{eqnarray}
\begin{eqnarray}
&& i\cos\theta\bigg(\frac{1}{6}Y_{\sharp\bar{1}}-\frac{1}{2}\Omega_{\sharp,\sharp\bar{1}}-\frac{1}{6}X_{\sharp\bar{1}p}^{\,\,\,\,\,\,\,\,\, p}+\frac{1}{18}X_{\sharp pqr}\epsilon^{pqr}\bigg)+\frac{1}{\sqrt{2}}\partial_{\sharp}\sin\theta
\nonumber\\
&+&\frac{1}{\sqrt{2}}\sin\theta\bigg(-\frac{1}{4}h_{\sharp}+\frac{1}{2}\Omega_{\sharp, p}^{\,\,\,\,\,\,\,p}-\frac{1}{2}\Omega_{\sharp,1\bar{1}}+\frac{1}{12}Y_p^{\,\,p}-\frac{1}{12}Y_{1\bar{1}}+\frac{1}{12}X_{1\bar{1}q}^{\,\,\,\,\,\,\,\,q}
\nonumber\\
&+&\frac{1}{24}X_{pq}^{\,\,\,\,\,pq}+\frac{1}{18}X_{\bar{1}pqr}\epsilon^{pqr}\bigg)=0\,,
\label{SING4}
\end{eqnarray}
\begin{eqnarray}
&& i\cos\theta\bigg(\frac{1}{6}Y_{\sharp}^{\,\,p}-\frac{1}{2}\Omega_{\sharp,\sharp}^{\,\,\,\,\,\,\,\,p}-\frac{1}{6}X_{\sharp1\bar{1}}^{\,\,\,\,\,\,\,\,\,p}-\frac{1}{6}X_{\sharp q}^{\,\,\,\,\,\,qp}-\frac{1}{6}X_{\sharp 1qr}\epsilon^{pqr}\bigg)
\nonumber\\
&+&\frac{1}{\sqrt{2}}\sin\theta\bigg(-\Omega_{\sharp,1}^{\,\,\,\,\,\,\,p}-\frac{1}{2}\Omega_{\sharp,qr}\epsilon^{pqr}-\frac{1}{6}Y_{1}^{\,\,p}-\frac{1}{12}Y_{qr}\epsilon^{pqr}
\nonumber\\
&+&\frac{1}{6}X_{1q}^{\,\,\,\,\,qp}+\frac{1}{12}X_{1\bar{1}qr}\epsilon^{pqr}-\frac{1}{36}X_{qrs}^{\,\,\,\,\,\,\,\,\,p}\epsilon^{qrs}\bigg)=0\,,
\label{jem}
\end{eqnarray}
\begin{eqnarray}
&& \frac{i}{\sqrt{2}}\cos\theta\bigg(\Omega_{\sharp,\bar{1}}^{\,\,\,\,\,\,\,\,p}-\frac{1}{6}Y_{\bar{1}}^{\,\,\,p}+\frac{1}{6}X_{\bar{1}q}^{\,\,\,\,\,qp}-\frac{1}{12}Y_{qr}\epsilon^{pqr}+\frac{1}{2}\Omega_{\sharp,qr}\epsilon^{pqr}
\nonumber\\
&-&\frac{1}{36}X_{qrs}^{\,\,\,\,\,\,\,\,\,p}\epsilon^{qrs}-\frac{1}{12}X_{1\bar{1}qr}\epsilon^{pqr}\bigg)+\sin\theta\bigg(-\frac{1}{6}Y_{\sharp}^{\,\,p}-\frac{1}{2}\Omega_{\sharp,\sharp}^{\,\,\,\,\,\,\,\,p}-\frac{1}{6}X_{\sharp1\bar{1}}^{\,\,\,\,\,\,\,\,\,p}
\nonumber\\
&+&\frac{1}{6}X_{\sharp t}^{\,\,\,\,\,tp}+\frac{1}{6}X_{\sharp\bar{1}qr}\epsilon^{pqr}\bigg)=0\,.
\label{ULTS22}
\end{eqnarray}

\appendix{$N=4$ $SU(3)$ KSE: 4-form $X$}
\label{appendix X}
The covariantized expression of the 4-form $X$ in the $SU(3)$ case is given by
\begin{eqnarray}
X&=&K\wedge U\wedge W\wedge C_1+K\wedge U\wedge C_2+K\wedge W\wedge C_3+U\wedge W\wedge C_4+K\wedge C_5
\nonumber \\
&+&U\wedge C_6+W\wedge C_7+C_8\,,
\label{ansatz}
\end{eqnarray}
where $C_1\in\Omega^1(\mathcal{S})$, $C_2,C_3,C_4\in\Omega^2(\mathcal{S})$, $C_5,C_6,C_7\in\Omega^3(\mathcal{S})$ and $C_8\in\Omega^4(\mathcal{S})$ can be expressed in terms of the $SU(3)$ gauge-invariant bilinears as follows
\begin{itemize}
\item{$C_8$}
\bea
C_8=\mathcal{F}_8\wedge\Re(\widehat{\chi})+G\wedge\widehat{\omega}+X^{(2,2)\textrm{TT}}\,,
\label{Y8full}
\eea
where $X^{(2,2)\textrm{TT}}$ is the traceless $(2,2)$ part of $X$, given by (\ref{TT}), 
\bea
(\mathcal{F}_8)_{i}:=-\frac{1}{2}\tan\theta\,\Theta_{i}+\frac{1}{4\cos\theta}(dU)_{l_1l_2}\Re(\widehat{\chi})^{l_1l_2}_{\,\,\,\,\,\,\,\,i}\,,
\label{Z8full}
\eea
\bea
G_{ij}:=\frac{1}{2}\cot\theta J_{[i}^{\,\,\,\,l}M_{j]l}\,,
\label{Cijfull}
\eea
\bea
M_{ij}:=-\widehat{\omega}^{l_1l_2}\nabla_{(i}\Re(\widehat{\chi})_{j)l_1l_2}\,,
\eea
\item{$C_7$}
\bea
C_7&=&\frac{1}{2(1+\cos^2\theta)}\bigg\{\Re(\widehat{\chi})\bigg(\frac{\cos^2\theta+2}{2}\tan\theta i_W\Theta-\frac{\cos^2\theta}{\sin\theta}  i_K\theta^{(9)}_{\Re(\widehat{\chi})}\bigg)
\nonumber \\
&+&\Im(\widehat{\chi})\bigg(-\cot\theta i_W\theta^{(9)}_{\Re(\widehat{\chi})}+\frac{\sin\theta}{2} i_K\Theta\bigg)\bigg\}+\mathcal{F}_7\wedge\widehat{\omega}
\nonumber\\
&+&\frac{1}{4}\bigg(\cot\theta(\mathcal{L}_Wg)_{i_1j}\Im(\widehat{\chi})^j_{\,\,i_2i_3}+\frac{1}{\sin\theta}(\mathcal{L}_Kg)_{i_1j}\Re(\widehat{\chi})^j_{\,\,i_2i_3}\bigg)e^{i_1i_2i_3}\,,
\label{Y7full}
\eea
\begin{eqnarray}
(\mathcal{F}_7)_{i}:&=&-\frac{1}{4\sin\theta}(\mathcal{L}_K\widehat{\omega})_{j_1j_2}\Re(\widehat{\chi})^{j_1j_2}_{\,\,\,\,\,\,\,\,\,\,i}-(\mathcal{L}_KU)_{i}
\nonumber \\
&-&\frac{1}{8}\cot\theta(\mathcal{L}_W\widehat{\omega})_{j_1j_2}\Im(\widehat{\chi})^{j_1j_2}_{\,\,\,\,\,\,\,\,\,\,i}\,,
\label{Z7full}
\end{eqnarray}
\item{$C_6$}
\begin{eqnarray}
C_6&=&\Re(\widehat{\chi})\bigg(\frac{1+\sin^2\theta}{2\sin(2\theta)}i_U\Theta\bigg)+\Im(\widehat{\chi})\bigg(-\frac{\cot\theta}{2(1+\cos^2\theta)}i_U\theta^{(9)}_{\Re(\widehat{\chi})}\bigg)
\nonumber\\
&+&\mathcal{F}_6\wedge\widehat{\omega}+\frac{1}{4}\cot\theta(\mathcal{L}_Ug)_{i_1j}\Im(\widehat{\chi})^j_{\,\,i_2i_3}e^{i_1i_2i_3}\,,
\label{Y6full}
\end{eqnarray}
\bea
(\mathcal{F}_6)_{i}:=\frac{1}{8}\cot\theta\bigg((dU)_{j_1j_2}\Re(\widehat{\chi})^{j_1j_2}_{\,\,\,\,\,\,\,\,\,\,i}-(\mathcal{L}_U\widehat{\omega})_{j_1j_2}\Im(\widehat{\chi})^{j_1j_2}_{\,\,\,\,\,\,\,\,\,\,i}\bigg)\,,
\label{Z6full}
\eea
\item{$C_5$}
\begin{eqnarray}
C_5&=&\frac{1}{2(1+\cos^2\theta)}\bigg\{\Re(\widehat{\chi})\bigg(\frac{\cos^2\theta+2}{2}\tan\theta i_K\Theta+\frac{\cos^2\theta}{\sin\theta}  i_W\theta^{(9)}_{\Re(\widehat{\chi})}\bigg)
\nonumber\\
&+&\Im(\widehat{\chi})\bigg(-\cot\theta i_K\theta^{(9)}_{\Re(\widehat{\chi})}-\frac{\sin\theta}{2} i_W\Theta\bigg)\bigg\}+\mathcal{F}_5\wedge\widehat{\omega}
\nonumber\\
&+&\frac{1}{4}\bigg(\cot\theta(\mathcal{L}_Kg)_{i_1j}\Im(\widehat{\chi})^j_{\,\,i_2i_3}-\frac{1}{\sin\theta}(\mathcal{L}_Wg)_{i_1j}\Re(\widehat{\chi})^j_{\,\,i_2i_3}\bigg)e^{i_1i_2i_3}\,,
\label{Y5full}
\end{eqnarray}
\bea
(\mathcal{F}_5)_{i}:=\frac{1}{4\sin\theta}(\mathcal{L}_W\widehat{\omega})_{j_1j_2}\Re(\widehat{\chi})^{j_1j_2}_{\,\,\,\,\,\,\,\,\,\,i}+(\mathcal{L}_WU)_{i}-\frac{\cot\theta}{8}(\mathcal{L}_K\widehat{\omega})_{j_1j_2}\Im(\widehat{\chi})^{j_1j_2}_{\,\,\,\,\,\,\,\,\,\,i}\,, \qquad
\label{Z5full}
\eea
\item{$C_4$}
\begin{eqnarray}
C_4&=&-\mathcal{L}_K\widehat{\omega}+\frac{1}{2}\bigg\{2J_{i_1}^{\,\,\,j}(\mathcal{L}_Kg)_{ji_2}+\bigg(\frac{4\cos^2\theta-3}{4\sin\theta}(\mathcal{L}_KU)_j-\frac{1}{4\sin\theta}[K,U]_j
\nonumber\\
&+&\frac{1}{2}\cot\theta J_j^{\,\,l}[W,U]_l\bigg)\Re(\widehat{\chi})^j_{\,\,i_1i_2}\bigg\}e^{i_1i_2}\,,
\label{Y4full}
\end{eqnarray}
\item{$C_3$}
\begin{eqnarray}
C_3&=&\mathcal{L}_U\widehat{\omega}+\frac{1}{2}\bigg\{-J_{i_1}^{\,\,\,j}\bigg(2(\mathcal{L}_Ug)_{ji_2}+(dU)_{i_2j}+\frac{1}{2\sin\theta} M_{i_2j}\bigg)
\nonumber \\
&+&\bigg(\sin\theta(\mathcal{L}_VV)_j+
+\frac{\cos^2\theta}{\sin\theta}(\mathcal{L}_KK)_j-\frac{1}{2}{\cot\theta} J_j^{\,\,l}[K,W]_l\bigg)\Re(\widehat{\chi})^j_{\,\,i_1i_2}\bigg\}e^{i_1i_2}
\nonumber\\
&+&\frac{1}{2}\widehat{\omega}\bigg(\frac{1}{2}i_U\theta^{(9)}_{\Re(\widehat{\chi})}-\star_9 d\star_9 U\bigg)\,,
\label{Y3full}
\end{eqnarray}
\item{$C_2$}
\begin{eqnarray}
C_2&=&-\mathcal{L}_W\widehat{\omega}+\frac{1}{2}\bigg\{2J_{i_1}^{\,\,\,j}(\mathcal{L}_Wg)_{ji_2}+\bigg(\frac{4\cos^2\theta-3}{4\sin\theta}(\mathcal{L}_WU)_j-\frac{1}{4\sin\theta}[W,U]_j
\nonumber\\
&-&\frac{1}{2}\cot\theta J_j^{\,\,\,l}[K,U]_l\bigg)\Re(\widehat{\chi})^j_{\,\,i_1i_2}\bigg\}e^{i_1i_2}\,,
\label{Y2full}
\end{eqnarray}
\item{$C_1$}
\begin{eqnarray}
C_1&=&\bigg\{J_i^{\,\,j}\bigg(\frac{1}{2}(\theta^{(9)}_{\Re(\widehat{\chi})})_{j}-\frac{3}{4}(\theta^{(9)}_{\widehat{\omega}})_{j}+(\mathcal{L}_KK)_j-\frac{1}{2}(\mathcal{L}_UU)_j\bigg)
\nonumber\\
&-&\frac{1}{8}\sin\theta(dU)_{l_1l_2}\Re(\widehat{\chi})^{l_1l_2}_{\,\,\,\,\,\,\,\,\,\,i}\bigg\}e^i-\cos\theta\mathcal{L}_KW\,.
\label{Y1full}
\end{eqnarray}
\end{itemize}

\appendix{Useful $Spin(7)$ and $G_2$ identities}

In this appendix we list some useful identities we have used extensively during our work.

The fundamental $Spin(7)$ 4-form $\tau$, defined by (\ref{tauuu}) satisfies the following identities
\begin{eqnarray}
\tau_{il_1l_2l_3}\tau^{jl_1l_2l_3}&=&42\delta_i^{\,\,j}
\nonumber\\
\tau_{i_1i_2l_1l_2}\tau^{j_1j_2l_1l_2}&=&-4\tau_{i_1i_2}^{\,\,\,\,\,\,\,\,\,l_1l_2}+12\delta^{[j_1}_{\,\,[i_1}\delta^{j_2]}_{\,\,i_2]}\,,
\nonumber\\
\tau_{i_1i_2i_3l}\tau^{j_1j_2j_3l}&=&-9\delta^{[j_1}_{\,\,[i_1}\tau_{i_2i_3]}^{\,\,\,\,\,\,\,\,\,\,\,j_2j_3]}+6\delta^{[j_1}_{\,\,[i_1}\delta^{j_2}_{\,\,i_2}\delta^{j_3]}_{\,\,i_3]}\,,
\label{identities}
\end{eqnarray}
where $i,j,...$ are 8-dimensional. The existence of a $Spin(7)$-structure 
on a 8-dimensional manifold $M^8$ determines a decomposition of the space 
$\Omega^k$ of the $k$-differential forms on $M^8$ into irreducible $Spin(7)$-representations $\Omega^k_{\textbf{n}}$ of dimension $\textbf{n}$, see \cite{fern} and also \cite{swal}.
$\Omega^1$ is irreducible; also, for $k>4$, $\Omega^k=\star_8\Omega^{8-k}$, thus we are left to decompose 
\begin{itemize}
\item{2-forms}
\bea
\Omega^2=\Omega^2_{\textbf{7}}\oplus\Omega^2_{\textbf{21}}\,,
\eea
where, given a $2$-form $\theta$
\begin{eqnarray}
\theta^{\textbf{7}}_{ij}&=&\frac{1}{4}\big(\theta_{ij}-\frac{1}{2}\tau_{ij}^{\,\,\,\,\,\,kl}\theta_{kl}\big)~,
\nonumber\\
\theta^{\textbf{21}}_{ij}&=&\frac{3}{4}\big(\theta_{ij}+\frac{1}{6}\tau_{ij}^{\,\,\,\,\,\,kl}\theta_{kl}\big)~.
\label{2formirr}
\end{eqnarray}
\item{3-forms}
\bea
\Omega^3=\Omega^3_{\textbf{8}}\oplus\Omega^3_{\textbf{48}}\,,
\eea
where, given a $3$-form $\lambda$
\begin{eqnarray}
\lambda^{\textbf{8}}_{ijk}&=&\frac{1}{7}\big(\lambda_{ijk}-\frac{3}{2}\tau^{mn}_{\,\,\,\,\,[ij}\lambda_{k]mn})\,,
\nonumber\\
\lambda^{\textbf{48}}_{ijk}&=&\frac{6}{7}\big(\lambda_{ijk}+\frac{1}{4}\tau^{mn}_{\,\,\,\,\,[ij}\lambda_{k]mn})\,.
\label{3formirr}
\end{eqnarray}
Notice that
\bea
\lambda_{j_1j_2j_3}\tau^{j_1j_2j_3}_{\,\,\,\,\,\,\,\,\,\,\,\,\,\,\,\,\,i}=\lambda^{\textbf{8}}_{j_1j_2j_3}\tau^{j_1j_2j_3}_{\,\,\,\,\,\,\,\,\,\,\,\,\,\,\,\,\,i}\,.
\eea
\item{4-forms}
\bea
\Omega^4=\Omega^4_{\textbf{1}}\oplus\Omega^4_{\textbf{7}}\oplus\Omega^4_{\textbf{27}}\oplus\Omega^4_{\textbf{35}}\,,
\eea
where, given a $4$-form $\xi$
\begin{eqnarray}
\xi^{\textbf{1}}_{i_1i_2i_3i_4}&=&\frac{1}{336}(\Pi^1 \xi)_{i_1i_2i_3i_4}\,,
\nonumber\\
\xi^{\textbf{7}}_{i_1i_2i_3i_4}&=&\frac{1}{8}\xi_{i_1i_2i_3i_4}+\frac{1}{224}(\Pi^1 \xi)_{i_1i_2i_3i_4}-\frac{3}{224}(\Pi ^2\xi)_{i_1i_2i_3i_4}+\frac{5}{168}(\Pi^3\xi)_{i_1i_2i_3i_4}\,,
\nonumber\\
\xi^{\textbf{27}}_{i_1i_2i_3i_4}&=&\frac{3}{8}\xi_{i_1i_2i_3i_4}-\frac{1}{224}(\Pi^1 \xi)_{i_1i_2i_3i_4}+\frac{15}{224}(\Pi^2 \xi)_{i_1i_2i_3i_4}+\frac{1}{56}(\Pi^3\xi)_{i_1i_2i_3i_4}\,,
\nonumber\\
\xi^{\textbf{35}}_{i_1i_2i_3i_4}&=&\frac{1}{2}\xi_{i_1i_2i_3i_4}-\frac{1}{336}(\Pi^1 \xi)_{i_1i_2i_3i_4}-\frac{3}{56}(\Pi^2\xi)_{i_1i_2i_3i_4}-\frac{1}{21}(\Pi^3\xi)_{i_1i_2i_3i_4}\,, \qquad 
\label{4form}
\end{eqnarray}
with
\begin{eqnarray}
(\Pi^1\xi)_{i_1i_2i_3i_4}&:=&\tau_{i_1i_2i_3i_4}\tau^{j_1j_2j_3j_4}\xi_{j_1j_2j_3j_4}\,,
\nonumber\\
(\Pi^2\xi)_{i_1i_2i_3i_4}&:=&\tau^{j_1j_2}_{\,\,\,\,[i_1i_2}\tau_{i_3i_4]}^{\,\,\,\,\,\,\,j_3j_4}\xi_{j_1j_2j_3j_4}\,,
\nonumber\\
(\Pi^3\xi)_{i_1i_2i_3i_4}&:=&\tau^{j_1}_{\,\,\,[i_1i_2i_3}\tau_{i_4]}^{\,\,\,j_2j_3j_4}\xi_{j_1j_2j_3j_4}\,.
\end{eqnarray}
\end{itemize}
The fundamental $G_2$ 3-form $\phi$, defined by (\ref{G2fundamental}), and its Hodge dual $\star_7\phi$ satisfy the following identities 
($i,j,..$ are 7-dimensional real $G_2$ indices)
\bea
\phi_{il_1l_2}\phi^{jl_1l_2}=6\delta_i^{\,\,j}\,,
\eea
\bea
\phi_{i_1i_2l}\phi^{j_1j_2l}=2\delta^{[j_1}_{\,\,[i_1}\delta^{j_2]}_{\,\,i_2]}-\psi_{i_1i_2}^{\,\,\,\,\,\,\,\,j_1j_2}\,,
\eea
\bea
\phi_{i}^{\,\,l_1l_2}(\star_7\phi)_{j_1j_2l_1l_2}=-4\phi_{ij_1j_2}\,,
\eea
\bea
\phi_{i_1i_2l}(\star_7\phi)^{j_1j_2j_3l}=6\delta^{[j_1}_{[i_1}\phi^{\,\,\,\,\,\,\,j_2j_3]}_{i_2]}\,.
\eea
\bea
(\star_7\phi)_{il_1l_2}(\star_7\phi)^{jl_1l_2}=24\delta_i^{\,\,j}\,,
\eea
\bea
(\star_7\phi)_{i_1i_2l_1l_2}(\star_7\phi)^{j_1j_2l_1l_2}=8\delta^{[j_1}_{\,\,[i_1}\delta^{j_2]}_{\,\,i_2]}-2(\star_7\phi)_{i_1i_2}^{\,\,\,\,\,\,\,\,\,\,\,j_1j_2}\,,
\eea
\bea
(\star_7\phi)_{i_1i_2i_3l}(\star_7\phi)^{j_1j_2j_3l}=6\delta^{[j_1}_{[i_1}\delta^{j_2}_{i_2}\delta^{j_3]}_{i_3]}-9\delta^{[j_1}_{[i_1}(\star_7\phi)_{i_2i_3]}^{\,\,\,\,\,\,\,\,\,\,\,j_2j_3]}-\phi_{i_1i_2i_3}(\star_7\phi)^{j_1j_2j_3}\,.
\eea
The existence of a $G_2$-structure on a 7-dimensional manifold $M^7$ determines a decomposition of the space $\Omega^k$ of the $k$-differential forms on $M^7$ into irreducible $G_2$-representations $\Omega^k_{\textbf{n}}$ of dimension $\textbf{n}$. The space $\Omega^k$ is irreducible if $k=1,6,7$ and $\Omega^k=\star_7\Omega^{7-k}$, thus we are left to decompose (see \cite{ferngray}):
\begin{itemize}
\item{2-forms}
\bea
\Omega^2=\Omega^2_{\textbf{7}}\oplus\Omega^2_{\textbf{14}}\,,
\label{2formsG2}
\eea
where, given a 2-form $\theta$
\begin{eqnarray}
\theta^{\textbf{7}}_{i_1i_2}&=&\frac{1}{3}\theta_{i_1i_2}-\frac{1}{6}\psi_{i_1i_2}^{\,\,\,\,\,\,\,\,\,j_1j_2}\theta_{j_1j_2}~,
\nonumber\\
\theta^{\textbf{14}}_{i_1i_2}&=&\frac{2}{3}\theta_{i_1i_2}+\frac{1}{6}\psi_{i_1i_2}^{\,\,\,\,\,\,\,\,\,j_1j_2}\theta_{j_1j_2} \ .
\end{eqnarray}
Notice that 
\bea
\theta_{j_1j_2}\phi^{j_1j_2}_{\,\,\,\,\,\,\,\,\,\,i}=\theta^{\textbf{7}}_{j_1j_2}\phi^{j_1j_2}_{\,\,\,\,\,\,\,\,\,\,i}\,.
\eea
\item{3-forms}
\bea
\Omega^3=\Omega^2_{\textbf{1}}\oplus\Omega^2_{\textbf{7}}\oplus\Omega^2_{\textbf{27}}\,,
\label{2formsG2}
\eea
where, given a 3-form $\lambda$
\begin{eqnarray}
\lambda^{\textbf{1}}_{i_1i_2i_3}&=&\frac{1}{42}\phi^{j_1j_2j_3}\lambda_{j_1j_2j_3}\phi_{i_1i_2i_3}\,,
\nonumber\\
\lambda^{\textbf{7}}_{i_1i_2i_3}&=&\frac{1}{4}\lambda_{i_1i_2i_3}-\frac{1}{24}\phi^{j_1j_2j_3}\lambda_{j_1j_2j_3}\phi_{i_1i_2i_3}-\frac{3}{8}\lambda_{j_1j_2[i_1}\psi_{i_2i_3]}^{\,\,\,\,\,\,\,\,\,\,\,j_1j_2}\,,
\nonumber\\
\lambda^{\textbf{27}}_{i_1i_2i_3}&=&\frac{3}{4}\lambda_{i_1i_2i_3}+\frac{1}{56}\phi^{j_1j_2j_3}\lambda_{j_1j_2j_3}\phi_{i_1i_2i_3}+\frac{3}{8}\lambda_{j_1j_2[i_1}\psi_{i_2i_3]}^{\,\,\,\,\,\,\,\,\,\,\,\,j_1j_2}\,.
\label{3formsG2}
\end{eqnarray}
Notice that 
\bea
\lambda_{j_1j_2j_3}\psi^{j_1j_2j_3}_{\,\,\,\,\,\,\,\,\,\,\,\,\,\,\,\,i}=\lambda^{\textbf{7}}_{j_1j_2j_3}\psi^{j_1j_2j_3}_{\,\,\,\,\,\,\,\,\,\,\,\,\,\,\,\,i}\,.
\eea
The 4-forms are dual to the 3-forms; nonetheless, let us state explicitly 
the decomposition of the 4-forms
\bea
\Omega^4=\Omega^4_{\textbf{1}}\oplus\Omega^4_{\textbf{7}}\oplus\Omega^4_{\textbf{27}}\,,
\label{2formsG2}
\eea
where, given a 4-form $\xi$
\begin{eqnarray}
\xi^{\textbf{1}}_{i_1i_2i_3i_4}&=&\frac{1}{168}\psi^{j_1j_2j_3j_4}\xi_{j_1j_2j_3j_4}\psi_{i_1i_2i_3j_4}\,,
\nonumber\\
\xi^{\textbf{7}}_{i_1i_2i_3i_4}&=&\frac{1}{4}\xi_{i_1i_2i_3i_4}-\frac{1}{96}\psi^{j_1j_2j_3j_4}\xi_{j_1j_2j_3j_4}\psi_{i_1i_2i_3j_4}-\frac{3}{4}\psi^{j_1j_2}_{\,\,\,\,\,\,\,\,\,[i_1i_2}\xi_{i_3i_4]j_1j_2}\,,
\nonumber\\
\xi^{\textbf{27}}_{i_1i_2i_3}&=&\frac{3}{4}\xi_{i_1i_2i_3i_4}+\frac{1}{224}\psi^{j_1j_2j_3j_4}\xi_{j_1j_2j_3j_4}\psi_{i_1i_2i_3j_4}+\frac{3}{4}\psi^{j_1j_2}_{\,\,\,\,\,\,\,\,\,[i_1i_2}\xi_{i_3i_4]j_1j_2}\,. \qquad
\label{4formsG2}
\end{eqnarray}
Notice that 
\bea
\xi_{ij_1j_2j_3}\phi^{j_1j_2j_3}=\xi^{\textbf{7}}_{ij_1j_2j_3}\phi^{j_1j_2j_3}\,.
\eea
\end{itemize}

\section*{Acknowledgments}

DF is partially supported by the STFC DTP Grant
ST/S505742. JG is supported by the STFC Consolidated Grant ST/L000490/1.
The authors would like to thank Martin Wolf for useful conversations.

\section*{Data Management}

No additional research data beyond the data presented and cited in this work are needed to validate the research findings in this work.

\end{document}